\documentclass[10pt,amssymb,superscriptaddress,showkeys,twocolumn,prl]{revtex4-2}
\usepackage{graphicx}
\usepackage{dcolumn}
\usepackage{bm}
\usepackage{changes}
\usepackage{amsmath,amssymb}
\makeatletter
\renewcommand{\section}{\@startsection{section}{1}{0mm}
{-\baselineskip}{0.5\baselineskip}{\bf\leftline}}
\makeatother
\makeatletter
\renewcommand{\subsection}{\@startsection{subsection}{1}{0mm}
{-\baselineskip}{0.5\baselineskip}{\bf\leftline}}

\begin{document}
\title{Observation of tunable topological polaritons in a cavity waveguide }% Force line breaks with \\

\author{Dong Zhao}
\thanks{These authors contributed equally}
\affiliation{Department of Electronic and Electrical Engineering, Southern University of Science and Technology, Shenzhen 518055, China.}
\author{Ziyao Wang}
\thanks{These authors contributed equally}
\affiliation{Department of Electronic and Electrical Engineering, Southern University of Science and Technology, Shenzhen 518055, China.}
\author{Linyun Yang}
\affiliation{Department of Electronic and Electrical Engineering, Southern University of Science and Technology, Shenzhen 518055, China.}
\author{Yuxin Zhong}
\affiliation{Department of Electronic and Electrical Engineering, Southern University of Science and Technology, Shenzhen 518055, China.}
\author{Xiang Xi}
\affiliation{Department of Electronic and Electrical Engineering, Southern University of Science and Technology, Shenzhen 518055, China.}
\author{Zhenxiao Zhu}
\affiliation{Department of Electronic and Electrical Engineering, Southern University of Science and Technology, Shenzhen 518055, China.}
\author{Maohua Gong}
\affiliation{Department of Electronic and Electrical Engineering, Southern University of Science
and Technology, Shenzhen 518055, China.}
\author{Qingan Tu}
\affiliation{Department of Electronic and Electrical Engineering, Southern University of Science
and Technology, Shenzhen 518055, China.}
\author{Yan Meng}
\email{mengy@sustech.edu.cn }
\affiliation{Department of Electronic and Electrical Engineering, Southern University of Science
and Technology, Shenzhen 518055, China.}
\author{Bei Yan}
\email{yanb3@sustech.edu.cn}
\affiliation{Department of Electronic and Electrical Engineering, Southern University of Science
and Technology, Shenzhen 518055, China.}
\author{Ce Shang}
\email{shang.ce@kaust.edu.sa}
\affiliation{King Abdullah University of Science and Technology (KAUST), Physical Science and Engineering Division (PSE), Thuwal 23955-6900, Saudi Arabia.}
\author{Zhen Gao}
\email{gaoz@sustech.edu.cn}
\affiliation{Department of Electronic and Electrical Engineering, Southern University of Science
and Technology, Shenzhen 518055, China.}
\affiliation{Laboratory of Optical Fiber and Cable Manufacture Technology, Southern University of Science and Technology, Shenzhen, Guangdong, China.}
\affiliation{Guangdong Key Laboratory of Integrated Optoelectronics Intellisense, Southern University of Science and Technology, Shenzhen, 518055, China.}
\date{\today}

\begin{abstract}
Topological polaritons characterized by light-matter interactions have become a pivotal platform in exploring new topological phases of matter. Recent theoretical advances unveiled a novel mechanism for tuning topological phases of polaritons by modifying the surrounding photonic environment (light-matter interactions) without altering the lattice structure. Here, by embedding a dimerized chain of microwave helical resonators (electric dipole emitters) in a metallic cavity waveguide, we report the pioneering observation of tunable topological phases of polaritons by varying the cavity width which governs the surrounding photonic environment and the strength of light-matter interactions. Moreover, we experimentally identified a new type of topological phase transition which includes three non-coincident critical points in the parameter space: the closure of the polaritonic bandgap, the transition of the Zak phase, and the hybridization of the topological edge states with the bulk states. These results reveal some remarkable and uncharted properties of topological matter when strongly coupled to light and provide an innovative design principle for tunable topological photonic devices.
\end{abstract}
\maketitle
Recent advances in topological photonics \cite{RevModPhys.82.3045, RevModPhys.83.1057, Khanikaev2012, Lu2014, RevModPhys.91.015006} and topological polaritonics \cite{PhysRevX.5.031001, PhysRevLett.114.116401, st2017lasing, klembt2018exciton, PhysRevLett.122.083902, Baranov2020, Liu_2020, wu2023higher, hu2023gate} have revolutionized our ability to manipulate light transcending the conventional boundaries of photonics. These breakthroughs have not only fostered a deeper understanding of the light-matter interactions at a fundamental level but also opened up entirely new avenues for various applications in diverse fields, such as topological waveguides \cite{ElHassan2019, chen2021photonic}, cavities \cite{bahari2017nonreciprocal, barczyk2022interplay}, lasers \cite{StJean2017, harari2018topological, bandres2018topological, yang2022topological}, integrated photonic circuits \cite{PhysRevX.5.021031, ma2019topological}, nonlinear \cite{kruk2019nonlinear, Smirnova2020}, and non-Hermitian photonics \cite{midya2018non, ElGanainy2018, parto2020non}. However, it is notoriously difficult, if not impossible, to manipulate the topological phases without modifying their lattice structures, since their topological invariants, such as the Zak phase in the one-dimensional (1D) \cite{PhysRevLett.62.2747} Su-Schrieffer-Heeger (SSH) model \cite{PhysRevLett.42.1698} and the Chern number in the two-dimensional (2D) topological photonic systems \cite{PhysRevLett.100.013904, wang2009observation}, are intrinsically determined by their lattice configurations. 

On the other hand, controlling light-matter interactions with cavities has played a fundamental role in modern science and technologies such as cavity quantum electrodynamics \cite{RevModPhys.73.565, walther2006cavity, mirhosseini2019cavity, owens2022chiral, lei2023many}, cavity magnonics \cite{PhysRevLett.120.057202, PhysRevLett.127.183202, PhysRevLett.129.123601}, and cavity plasmonics \cite{chanda2011coupling, hugall2018plasmonic, RevModPhys.94.025004}. More interestingly, recent theoretical studies \cite{ PhysRevLett.123.217401, mann2018manipulating, mann2020tunable, downing2020polaritonic} reveal that cavity-controlled light-matter interactions can tune the topological phases of polaritons in a cavity waveguide by modulating the surrounding photonic environment without changing the lattice structure. This novel universal mechanism has led to the theoretical discoveries of many previously unexplored topological phenomena, such as the breakdown of bulk-boundary correspondence \cite{PhysRevLett.123.217401}, the manipulation of type-I and type-II Dirac polaritons \cite{mann2018manipulating},  and the tunable pseudo-magnetic fields \cite{mann2020tunable}. However, so far the experimental observation of tunable topological polaritons in a cavity waveguide remains elusive. 

Here, by embedding a 1D dimerized chain of microwave helical resonators (MHRs) in a metallic cavity waveguide, we report the first experimental observation of tunable topological polaritons by modifying only the surrounding photonic environment (light-matter interactions) without altering the lattice configuration. We experimentally demonstrate that the intrinsic band topology (Zak phase) and polaritonic band structure of the composite structure can be fundamentally tuned by changing the cavity waveguide width. Moreover, we experimentally identify three non-coincident critical points in the parameter space: when the polaritonic bandgap closes, when the Zak phase changes from nontrivial to trivial, and when the topological edge states begin to hybridize with the bulk states, verifying a new type of topological phase transition that includes three different critical transition points \cite{PhysRevLett.123.217401, PhysRevLett.118.166803}. 

\begin{figure}[t]
\centering
\includegraphics[width=1\columnwidth]{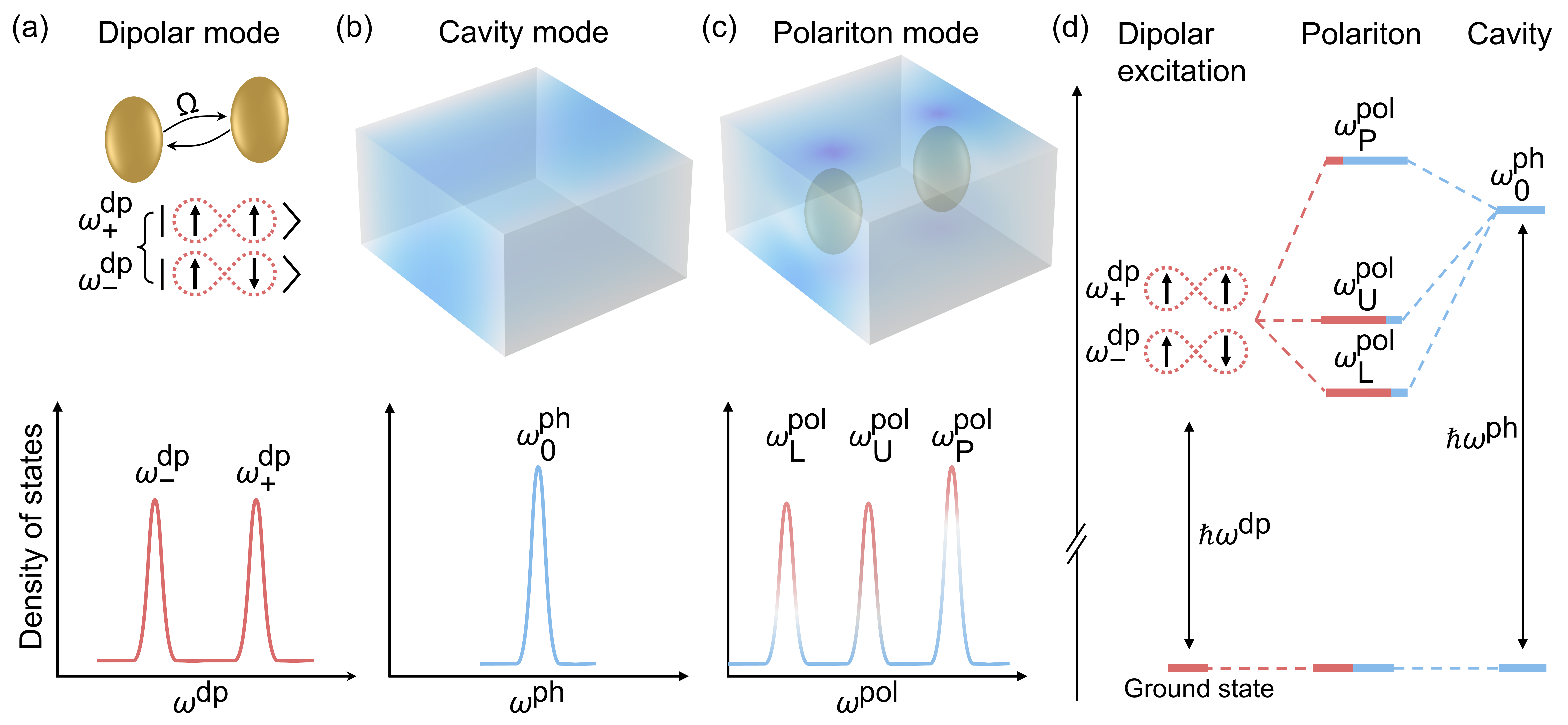}
\caption{Schematic illustration of light-matter interactions within a three-level model. (a) Two dipolar meta-atoms (upper panel). The Coulomb interaction ($\Omega$) leads to the splitting of two coupled dipole modes with eigenfrequencies $\omega_{-}^{\mathrm{dp}}$ (antisymmetric mode) and $\omega_{+}^{\mathrm{dp}}$ (symmetric mode). The density of states (DOS) for these modes displays two resonance peaks near their respective eigenfrequency regimes (lower panel). (b) The photonic cavity mode (upper panel) induces a single peak near its fundamental resonance eigenfrequency $\omega_{0}^{\mathrm{ph}}$. (c) A three-level model constructed by two dipolar meta-atoms embedded in the photonic cavity (upper panel). Mode hybridization produces three polaritonic modes with eigenfrequencies $\omega_{\mathrm{L}}^{\mathrm{pol}}$, $\omega_{\mathrm{U}}^{\mathrm{pol}}$, and $\omega_{\mathrm{P}}^{\mathrm{pol}}$ (lower panel).(d) Mode hybridization of a three-level system comprising two coupled dipole modes and a cavity mode. The relative occupations of dipolar and cavity modes are qualitatively indicated by red and blue hues, respectively. }
\label{fig1}
\end{figure}

To elucidate the polaritons formation in a metallic cavity waveguide, we first employ a conceptual framework of the light-matter interactions of two dipolar meta-atoms embedded within a photonic cavity and depict a three-level system comprising two coupled dipolar modes and a photonic cavity mode. As illustrated in the upper panel of Fig.\ \ref{fig1}(a), two dipoles are indicated by a pair of golden meta-atoms with the coupling characterized by the interaction strength $\Omega$, which induces the formation of two coupled dipolar modes: an antisymmetric mode with a lower eigenfrequency $\omega_{-}^{\mathrm{dp}}$, and a symmetric mode with a higher eigenfrequency $\omega_{+}^{\mathrm{dp}}$. The density of states for these coupled modes is depicted in the lower panel of Fig. \ref{fig1}(a), revealing two resonance peaks. The photonic cavity mode is shown in the upper panel Fig. \ref{fig1}(b), displaying a fundamental resonance peak near the eigenfrequency $\omega_{0}^{\mathrm{ph}}$ (lower panel). The embedding of the dipolar meta-atoms within a cavity, as illustrated in Fig. \ref{fig1}(c), results in collective strong light-matter interaction according to cavity/waveguide quantum electrodynamics \cite{RevModPhys.73.565, walther2006cavity, RevModPhys.95.015002}, producing hybrid polaritonic modes with the resonance frequencies of $\omega_{\mathrm{L}}^{\mathrm{pol}}$, $\omega_{\mathrm{U}}^{\mathrm{pol}}$, and $\omega_{\mathrm{P}}^{\mathrm{pol}}$, respectively. In Fig. \ref{fig1}(d), the manifestation of mode hybridization in this three-level system is depicted with three polaritonic modes, providing insight of the interplay between the meta-atoms and the cavity. The schematic representation employs a color scheme to delineate the relative occupations of dipolar and cavity modes within the system. The red hues indicate the predominant impact of the dipolar modes, whereas the blue hues signify the impact of the cavity modes. Two of the polaritonic modes $\omega_{L, U}^{\mathrm{pol}}$ exhibit lower eigenfrequencies stemming from their 'original' dipolar states $\omega_{\pm}^{\mathrm{dp}}$, while the third polaritonic mode, influenced by cavity modes, displays a higher eigenfrequency than the 'original' cavity modes \cite{Sturges_2020}.

To demonstrate the tunable topological phases of polaritons, we embed a dimerized chain of MHRs in a metallic cavity waveguide with two air gaps ($1\ \mathrm{mm}$) separating the MRHs chain from the upper and lower metallic plates (the air gaps are filled by a foam spacer), as schematically shown in Fig. 2(a). The metallic cavity waveguide has a fixed height of $L_z = 24 \ \mathrm{mm}$, while its width $L_y$ is tunable to modulate the surrounding photonic environment and light-matter-interaction strength. Each unit cell of the 1D dimerized chain (white cubic frame) contains two MHRs and has a lattice constant of $d = 40\ \mathrm{mm}$. The alternating center-to-center distances between two neighboring MHRs are $d_1 = 0.575d$ and $d_2 = 0.425d$, respectively. A single MHR is shown in Fig. 2(b) with dimensions: copper wire diameter $2r =2\ \mathrm{mm}$, helix diameter $2R = 15\ \mathrm{mm}$, helix height $h = 22\ \mathrm{mm}$, axial intercept $l = 5\ \mathrm{mm}$, and 4 turns. The 1D dimerized chain of MHRs supports collective dipolar (dp) excitations (oscillating electric dipoles), which can be modeled as a prototypical 1D SSH model in the microwave regime \cite{PhysRevB.95.125426}:
\begin{equation}
\mathcal{H}_{\mathrm{dp}}=\left(\begin{array}{cc}
\omega_{0} & \Omega g_{k_x} \\
\Omega g^*_{k_x} & \omega_{0}
\end{array}\right),    
\end{equation}
where $\omega_0$ is the resonance frequency of dipole, $\Omega=\omega_0(a / d)^3 / 2$ is the coupling constant, $a$ is a length scale characterizing the strength of the dipolar excitations, and $g_{k_x}=$ $\left(d / d_1\right)^3+\left(d / d_2\right)^3 e^{-i k_x d}$ is a function of wave vector along $x$ direction $k_x$. The resulting band structure for the dipolar excitations is given as $\omega_{\pm}^{\mathrm{dp}}=\omega_0 \pm \Omega|g_{k_x}|$, which shows a symmetric feature between the in-phase dipole momentums $\omega_{+}^{\mathrm{dp}}$ and out-of-phase dipole momentums $\omega_{-}^{\mathrm{dp}}$ with a bandgap of $2 \Omega|g_{k_x}|$.

When embedded within a metallic cavity waveguide, the dipolar excitations can couple with the fundamental photonic (ph) waveguide modes characterized by the dispersion relation of $\omega_{{k_x}}^{\mathrm{ph}}=c_0 \sqrt{k_x^2+\left(\pi / L_y\right)^2}$, where $c_0$ represents the speed of light in vacuum. This composite structure induces strong light-matter interactions between two dipolar modes and a photonic waveguide mode, upgrading the typical SU(2) SSH model (1D dimerized chain of MHRs) to a SU(3) polaritonic model (1D dimerized chain of MHRs embedded in a cavity waveguide) \cite{PhysRevLett.123.217401}:
\begin{equation}
\mathcal{H}_{\mathrm{pol}}=\left(\begin{array}{ccc}
\omega_0 & \Omega g_{k_x} & {\mathrm{i}} \xi_{k_x} e^{-{\mathrm{i}} \chi_{k_x}} \\
\Omega g_{k_x}^* & \omega_0 & {\mathrm{i}} \xi_{k_x} e^{{\mathrm{i}} \chi_{k_x}} \\
-{\mathrm{i}} \xi_{k_x} e^{{\mathrm{i}} \chi_{k_x}} & -{\mathrm{i}} \xi_{k_x} e^{-{\mathrm{i}} \chi_{k_x}} & \omega_{k_x}^{\mathrm{ph}}
\end{array}\right),
\end{equation}
where $\xi_{k_x}=\left(2 \pi a^3 \omega_0^3 / d L_y L_z \omega_{k_x}^{\mathrm{ph}}\right)^{1 / 2}$ indicates the strength of light-matter interactions, and $\chi_{k_x}={k_x} d_1 / 2$ stems from the phase difference between two inequivalent lattice sites within a unit cell.

We use the COMSOL Multiphysics RF Module to solve the phase diagram of the polaritonic band structures $\omega^{\mathrm {pol}}_{j=\{L, U, P\}}$ as a function of $L_y$ and $k_x$, as shown in Fig. 2(c). For $L_y<1.3d$, the polaritonic bands $\omega^{\mathrm {pol}}_{L, U}$ are smoothly deformed from the dipolar bands $\omega_{\pm}^{\mathrm{dp}}$ due to the negligible weak light-matter interactions $\xi_{k_x}$; for $1.3d <L_y<2.6d$, the center region (near $k_x = 0$) of $\omega^{\mathrm {pol}}_{U}$ descends via the increase of $\xi_{k_x}$ while $\omega^{\mathrm {pol}}_{L}$ remains almost unchanged, and the polaritonic bandgap closes at $L_y =2.4d$; for $L_y >2.6d$, $\omega^{\mathrm {pol}}_{L, U}$ display an anti-crossing phenomenon, pushing down the center region of $\omega^{\mathrm {pol}}_{L}$ while keeping $\omega^{\mathrm {pol}}_{U}$ almost unchanged. We select four critical cavity waveguide widths ($L_y = 1.1d$, $2.4d$, $2.6d$, and $3.0d$, respectively) and plot their simulated band structures in Figs. 2(d)-(g) with the z-component of electric field ($E_z$) distributions of the eigenmodes at $k_x = 0$ $(A, B, C)$ and $k_x = \pi / d$ $(D)$  shown in the insets. 

\begin{figure}[t]
\centering
\includegraphics[width=1\columnwidth]{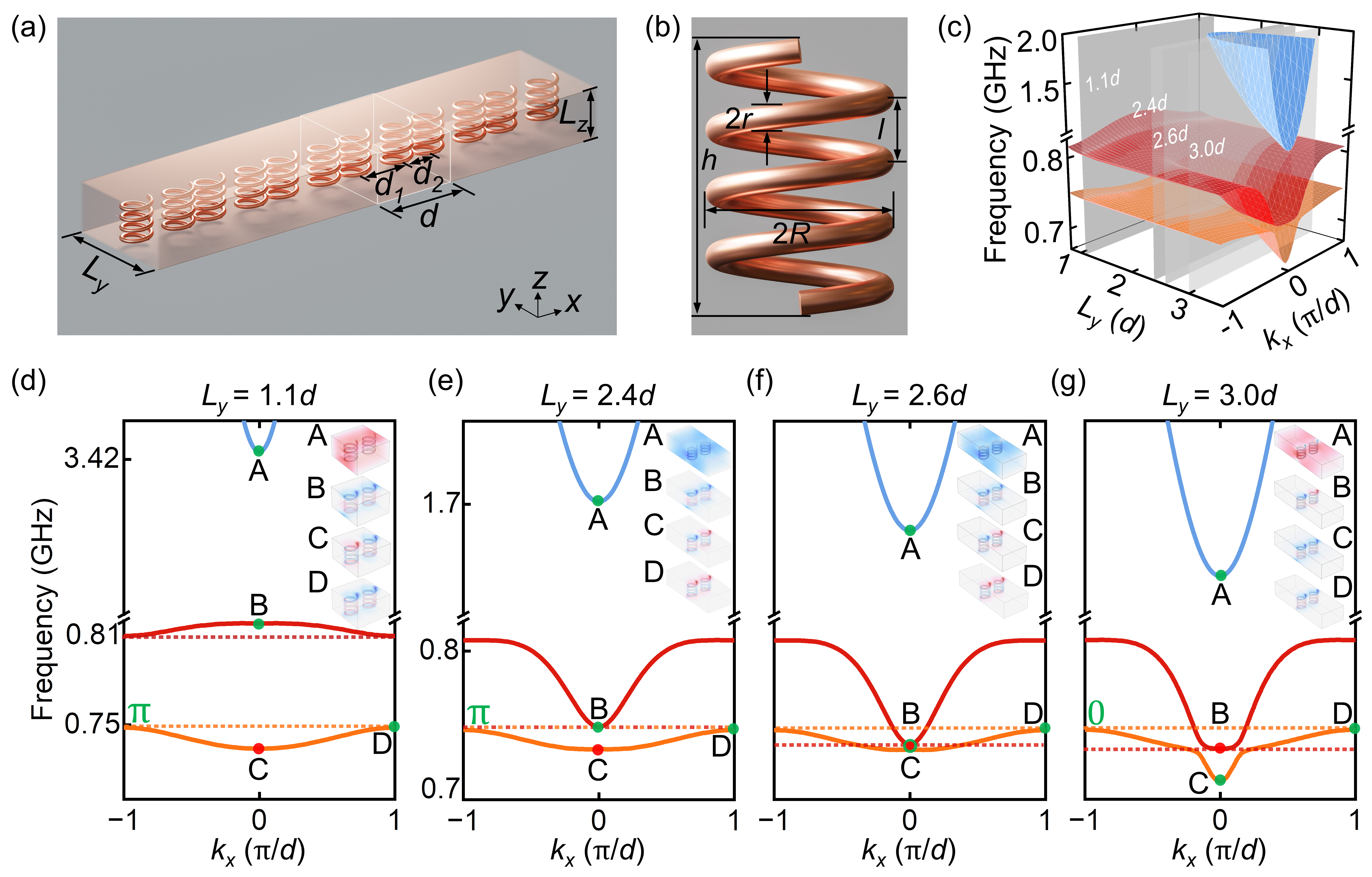}
\caption{Tunable topological phases of polaritons. (a) Schematic of the composite structure consisting of a 1D dimerized chain of MHRs embedded in a metallic cavity waveguide. (b) A single MHR is achieved by a copper wire with a diameter of $2r$, a helix diameter of $2R$, a helix height of $h$, and a helix axial intercept of $l$. (c) Simulated band structure diagram of the composite structure as a function of cavity waveguide width $L_y$ and wavevector $k_x$. (d)-(g) Simulated band structures for different cavity waveguide widths $L_y = 1.1d$ (d), $2.4d$ (e), $2.6d$ (f), and $3.0d$ (g), respectively. Green letters $\pi$ and 0 represent the Zak phases of the lowest polaritonic band ($\omega_\mathrm{L}^{\mathrm{pol}}$). Insets show the z-component of electric field ($E_z$) distributions of the eigenmodes at $k_x = 0$ $(A, B, C)$ and $k_x = \pi/d$ $(D)$. }
\label{fig2}
\end{figure}

Notably, the change of surrounding photonic environments (light-matter-interaction strengths) not only modulates the band structures but also tunes the topological phases of polaritons. The topological invariant for the $j$-th band is characterized by the Zak phase \cite{PhysRevX.4.021017}: 
\begin{equation} \label{S3}
\theta^{\mathrm{Zak}}_j=\mathrm{i} \int_{-\pi / d}^{+\pi / d} d {k_x}\left\langle\psi_{{k_x} ,j}\left|\partial_{k_x}\right| \psi_{{k_x} ,j}\right\rangle,
\end{equation}
which is quantized by 0 (topological trivial phase) or $\pi$ (topological nontrivial phase), where $\psi_{{k_x},j}$ is the periodic part of the eigenstates. A discretized form of Eq.\ \ref{S3} is defined by
\begin{equation} \label{S4}
  \theta _j^{{\mathrm{Zak}}} =  - {\mathop{\rm Im}\nolimits} \sum\limits_{n = 1}^{N - 1} {\ln } \left\langle {{\psi _{{k_{x_n},j}}}|{\psi _{{k_{x_{n + 1}},j}}}} \right\rangle ,  
\end{equation} 
where $N$ is the number of divided parts of $k_x$, $\psi_{j, {k_{x_n}}}$ is the discretized form of $\psi_{k_x, j}$ for a given momentum $k_{x_n}$. For the transverse magnetic modes corresponding to the electric field component $E_z$, $\psi_{k_{x_n}, j}$ can be numerically extracted from the equation: $\psi_{k_{x_n}, j}= E_{z; k_{x_n}, j}e^ {-\mathrm{i} k_x x}$, where $E_{z; k_{x_n}, j}$ represents the normalized $E_z$ in one unit cell. The Zak phases of $\omega^{\mathrm {pol}}_{L}$ versus $L_y$ are presented as green letters in Figs. \ref{fig2}(d)-(g), indicating the topological non-trivial (trivial) phases associated with $\pi(0)$ for $L_y<2.6 d\left(L_y>2.6 d\right)$. Additionally, the Zak phase can also be qualitatively determined by the symmetry of the eigenmodes at $k_x=0$ (mode $C$) and $k_x=\pi / d$ (mode $D$). If they exhibit different (same) mode symmetries, the Zak phase is $\pi$ (0) \cite{PhysRevX.4.021017, xiao2015geometric}, as supported by the calculated results.

\begin{figure}[t]
\centering
\includegraphics[width=1\columnwidth]{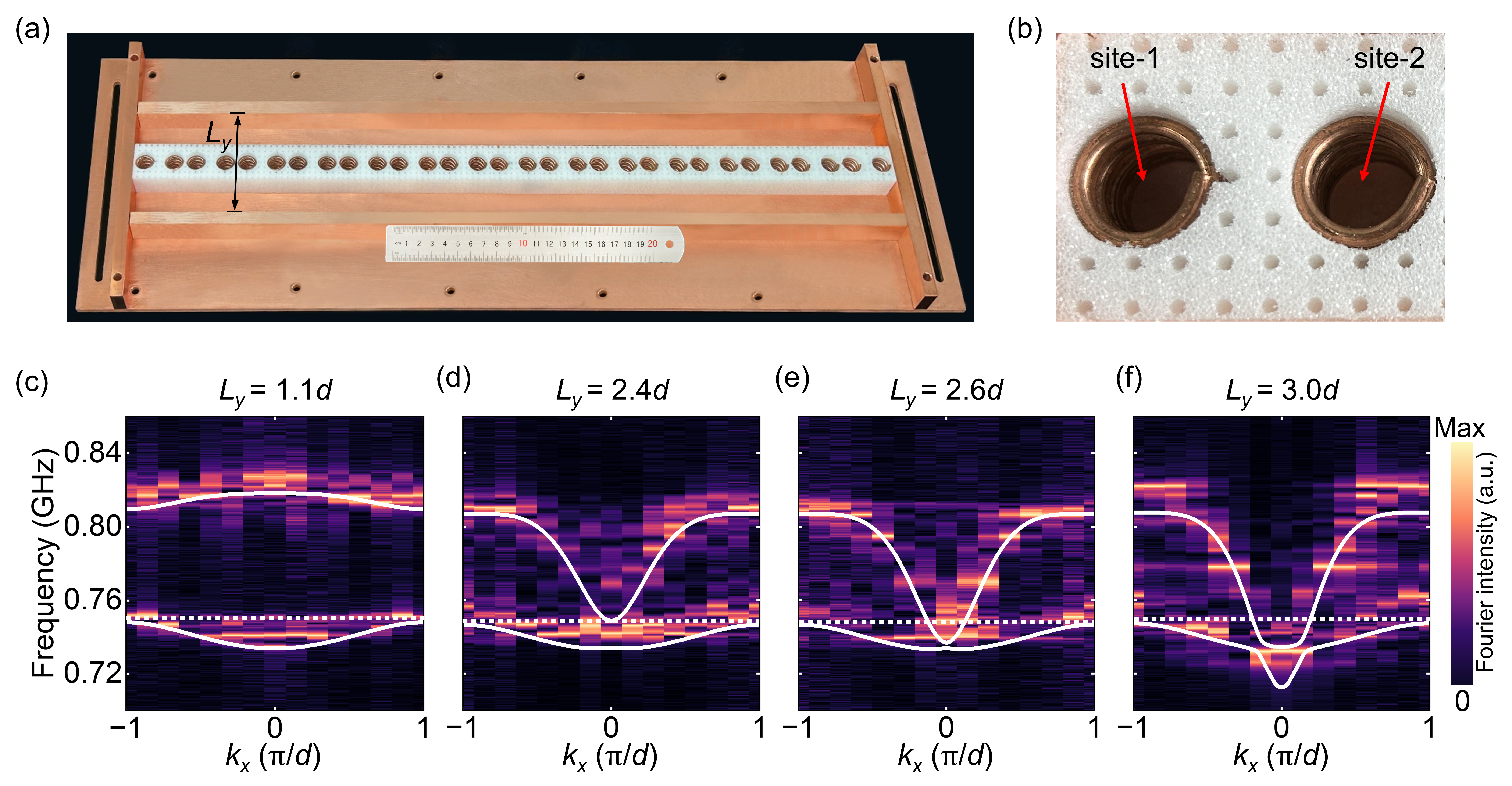}
\caption{Experimental observation of tunable polaritonic band structures by varying the cavity waveguide width $L_y$. (a) Photo of the fabricated sample consisting of a 1D dimerized chain of 30 MHRs embedded in a metallic cavity waveguide. (b) A unit cell of the 1D dimerized chain of MHRs. (c)-(f) Measured (color maps) and simulated (white solid lines) polaritonic band structures of the sample with different cavity waveguide widths $L_y$. The white dashed lines represent the maximum frequency of the lowest polaritonic band ($\omega_\mathrm{L}^{\mathrm{pol}}$).}
\label{fig3}
\end{figure}

To experimentally observe the tunable topological phases of polaritons, we fabricate a sample comprising a dimerized chain of 30 MHRs embedded in a metallic cavity waveguide, as shown in Fig.\ \ref{fig3}(a), where the upper metallic plate has been removed to see the inner structure. Each unit cell comprises two MHRs that are labeled site-1 and site-2, respectively, as shown in Fig.\ \ref{fig3}(b).
The experiment setup consists of a vector network analyzer (Agilent 5232A) and two electric monopole antennas, one as a point source to excite the composite structure and the other to measure the $E_z$ field distributions. We conduct Fourier transformation (FFT) to the measured electric field distributions to extract the measured polaritonic band structures (color maps), which agree well with the simulated results (white solid lines), as shown in Figs.\ \ref{fig3}(c)-3(f).  These experimental results unambiguously verify that the polaritonic band structures can be modified by only structuring the surrounding photonic environment (light-matter interactions). 

\begin{figure}[t]
\centering
\includegraphics[width=1\columnwidth]{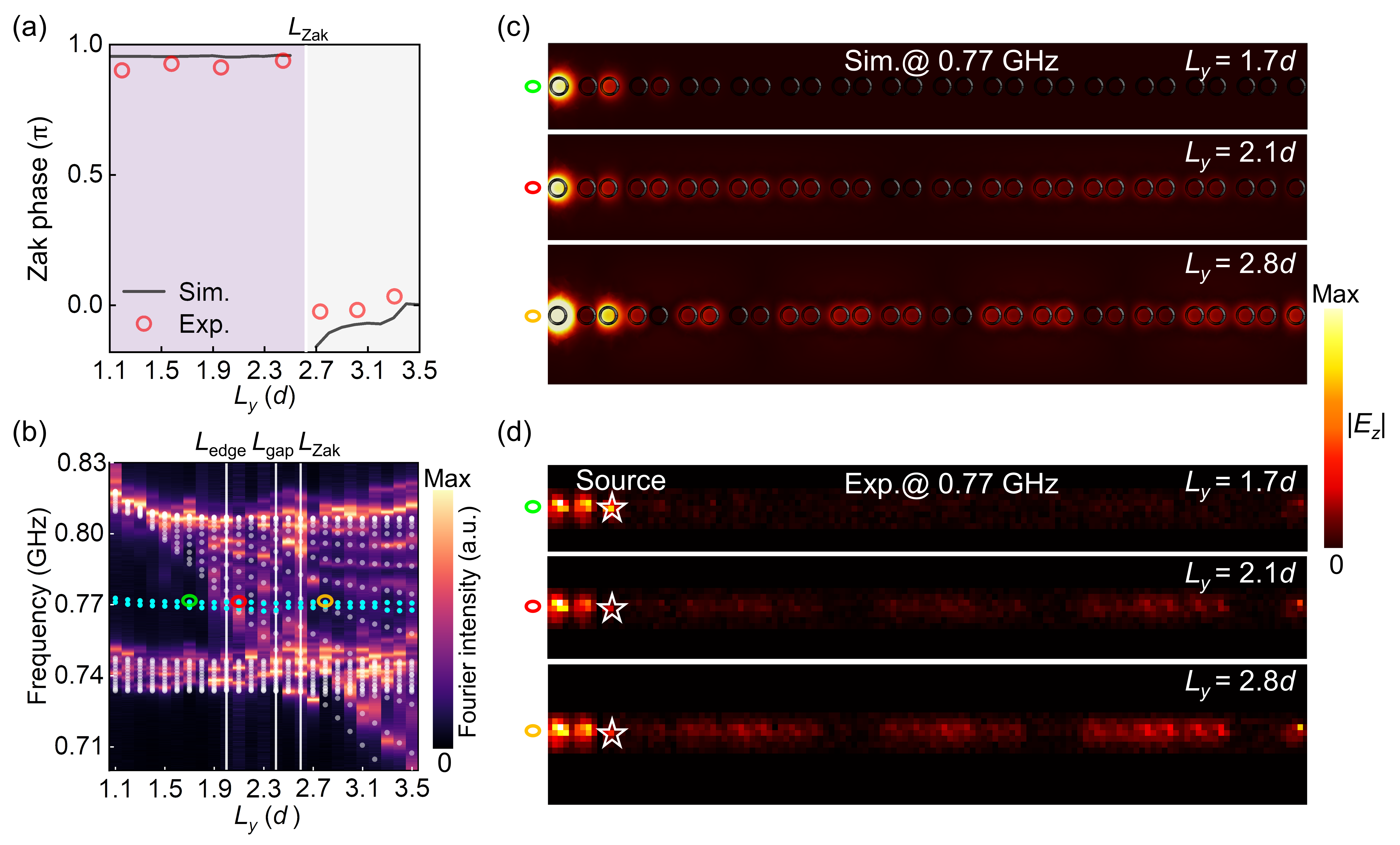}
\caption{Experimental observation of tunable topological phases of polaritons by changing the cavity waveguide width $L_y$. (a) Measured (red circle) and simulated (black lines) Zak phases of $\omega_\mathrm{L}^{\mathrm{pol}}$ as a function of  $L_y$. (b) Measured (color map) and simulated (white dots) eigenfrequency diagram of a 1D finite dimerized chain of 30 MHRs embedded in a cavity waveguide as a function of $L_y$. The simulated topological edge states are represented by cyan dots. Simulated (c) and measured (d) electric field distributions of the topological edge states at 0.77 GHz with $L_y = 1.7d$ (green ellipse), $2.1d$ (red ellipse), and $2.8d$ (brown ellipse), respectively. The white stars represent the point source. }
\label{fig4}
\end{figure}

To experimentally explore the topological phase transition and the bulk-boundary correspondence, we adopt a tight-binding approximation method to obtain the discrete eigenvectors using the formula $\psi_{k_{x_n}, \mathrm{L}}=(F_{k_{x_n}, \mathrm{L}}^1, F_{k_{x_n}, \mathrm{L}}^2)$, where $F_{k_{x_n}, \mathrm{L}}^1 (F_{k_{x_n}, \mathrm{L}}^2)$ corresponds to the amplitude peaks of the FFT spectra of $\omega_\mathrm{L}^{\mathrm{pol}}$ that can be extracted from the measured $E_z$ at a single sublattice site-1 (site-2) in each unit cell \cite{PhysRevLett.128.116803}. With the experimentally extracted $\psi_{k_{x_n}, \mathrm{L}}$, the Zak phase of $\omega_\mathrm{L}^{\mathrm{pol}}$ band can be obtained according to Eq.\ \ref{S4}. As shown in Fig.\ \ref{fig4}(a), the measured (red circles) and simulated (black line) Zak phases of $\omega_\mathrm{L}^{\mathrm{pol}}$ are plotted as a function of $L_y$. It can be observed that when we increase  $L_y$ from $1.1d$ to $3.5d$, the Zak phases evolve from being nontrivial ($\pi$) to trivial (0) with a topological phase transition point around $L_y = 2.6d$.

Finally, we investigate the evolution of the topological edge states versus $L_y$. Figure \ref{fig4}(b) shows the measured (color maps) and the simulated [white (cyan) dots represent the bulk (edge) states] polaritonic eigenfrequencies as a function of $L_y$. For $L_y=1.1 d$, we observe SSH-model-like band spectra due to negligible weak light-matter interactions, corresponding to Fig. \ref{fig2}(d) and Fig. \ref{fig3}(c). Two topological edge states (cyan dots) in the bandgap are localized at two ends of the MRHs chain. As $L_y$ increases from $1.1d$ to $3.5d$, we observe that the eigenfrequency spectra of $\omega_\mathrm{U}^{\mathrm{pol}}$ gradually approach those of $\omega_\mathrm{L}^{\mathrm{pol}}$, while the topological edge states remain almost unchanged. Consequently, the topological edge states merge into the bulk band at $L_{\mathrm{edge}} = 2.0d$ (left vertical white line), causing some delocalization of the topological edge states. This merging occurs before the polaritonic bandgap closes at $L_{\mathrm{gap}} = 2.4d$ (middle vertical white line) and the Zak phase changes at $L_{\mathrm{Zak}} = 2.6d$ (right vertical white line), which is totally different from the standard SSH model or even its extended version that involves beyond the nearest-neighbor couplings \cite{PhysRevB.99.035146}.
Now we show how the topological edge states evolve as we continuously increase $L_y$. We plot the simulated and measured $E_z$ field distributions of the topological edge states at 0.77 GHz for $L_y = 1.7d$ (green ellipse), $2.1d$ (red ellipse), and $2.8d$ (brown ellipse) in Fig.\ \ref{fig4}(c) and Fig.\ \ref{fig4}(d), respectively. We observe that before merging into the bulk band (${L_y < L_{\mathrm{edge}}}$), the topological edge states are located within the polaritonic bandgap and exhibit strong field localization at the edge of the 1D MHRs chain. However, when they merge into the bulk band (${L_y} > L_{\mathrm{edge}}$), the topological edge states begin to hybridize with the bulk states, resulting in novel mixed states (bound states in the continuum) with strong field localization near the edge and extended field distribution within the bulk simultaneously \cite{PhysRevLett.123.217401, PhysRevLett.118.166803}.

In conclusion, we have experimentally observed tunable topological phases of polaritons for the first time in a cavity-embedded dimerized chain of MHRs. We formulated a three-level system to elucidate the formation of topological polaritons.  Notably, we demonstrated that both the topological polaritonic band structures and topological invariant (Zak phase) can be tuned by modifying the surrounding photonic environment without altering the lattice structure. Furthermore, we experimentally identified a new type of topological phase transition which includes three non-coincident critical points in the parameter space: the closure of the polaritonic bandgap, the transition of the Zak phase, and the merging of the topological edge states with the bulk states. This work not only demonstrates a novel mechanism for tuning topological phases by modifying the surrounding photonic environment without altering the lattice structure but also establishes an ideal photonic platform to explore exotic topological physical phenomena emerging from light-matter interactions beyond the paradigm of conventional tight-binding physics. 

\begin{acknowledgements}
Z.G. acknowledges funding from the National Natural Science Foundation of China under Grant Nos. 62375118, 6231101016, and 12104211, Shenzhen Science and Technology Innovation Commission under grant no. 20220815111105001, and SUSTech under Grant Nos. Y01236148 and Y01236248. Y.M. acknowledges the support from the National Natural Science Foundation of China under Grant No. 12304484.
\end{acknowledgements}
\providecommand{\noopsort}[1]{}\providecommand{\singleletter}[1]{#1}%


\begin{thebibliography}{56}%
\makeatletter
\providecommand \@ifxundefined [1]{%
 \@ifx{#1\undefined}
}%
\providecommand \@ifnum [1]{%
 \ifnum #1\expandafter \@firstoftwo
 \else \expandafter \@secondoftwo
 \fi
}%
\providecommand \@ifx [1]{%
 \ifx #1\expandafter \@firstoftwo
 \else \expandafter \@secondoftwo
 \fi
}%
\providecommand \natexlab [1]{#1}%
\providecommand \enquote  [1]{``#1''}%
\providecommand \bibnamefont  [1]{#1}%
\providecommand \bibfnamefont [1]{#1}%
\providecommand \citenamefont [1]{#1}%
\providecommand \href@noop [0]{\@secondoftwo}%
\providecommand \href [0]{\begingroup \@sanitize@url \@href}%
\providecommand \@href[1]{\@@startlink{#1}\@@href}%
\providecommand \@@href[1]{\endgroup#1\@@endlink}%
\providecommand \@sanitize@url [0]{\catcode `\\12\catcode `\$12\catcode `\&12\catcode `\#12\catcode `\^12\catcode `\_12\catcode `\%12\relax}%
\providecommand \@@startlink[1]{}%
\providecommand \@@endlink[0]{}%
\providecommand \url  [0]{\begingroup\@sanitize@url \@url }%
\providecommand \@url [1]{\endgroup\@href {#1}{\urlprefix }}%
\providecommand \urlprefix  [0]{URL }%
\providecommand \Eprint [0]{\href }%
\providecommand \doibase [0]{https://doi.org/}%
\providecommand \selectlanguage [0]{\@gobble}%
\providecommand \bibinfo  [0]{\@secondoftwo}%
\providecommand \bibfield  [0]{\@secondoftwo}%
\providecommand \translation [1]{[#1]}%
\providecommand \BibitemOpen [0]{}%
\providecommand \bibitemStop [0]{}%
\providecommand \bibitemNoStop [0]{.\EOS\space}%
\providecommand \EOS [0]{\spacefactor3000\relax}%
\providecommand \BibitemShut  [1]{\csname bibitem#1\endcsname}%
\let\auto@bib@innerbib\@empty
%</preamble>
\bibitem [{\citenamefont {Hasan}\ and\ \citenamefont {Kane}(2010)}]{RevModPhys.82.3045}%
  \BibitemOpen
  \bibfield  {author} {\bibinfo {author} {\bibfnamefont {M.~Z.}\ \bibnamefont {Hasan}}\ and\ \bibinfo {author} {\bibfnamefont {C.~L.}\ \bibnamefont {Kane}},\ }\bibfield  {title} {\bibinfo {title} {Colloquium: Topological insulators},\ }\href {https://doi.org/10.1103/RevModPhys.82.3045} {\bibfield  {journal} {\bibinfo  {journal} {Rev. Mod. Phys.}\ }\textbf {\bibinfo {volume} {82}},\ \bibinfo {pages} {3045} (\bibinfo {year} {2010})}\BibitemShut {NoStop}%
\bibitem [{\citenamefont {Qi}\ and\ \citenamefont {Zhang}(2011)}]{RevModPhys.83.1057}%
  \BibitemOpen
  \bibfield  {author} {\bibinfo {author} {\bibfnamefont {X.-L.}\ \bibnamefont {Qi}}\ and\ \bibinfo {author} {\bibfnamefont {S.-C.}\ \bibnamefont {Zhang}},\ }\bibfield  {title} {\bibinfo {title} {Topological insulators and superconductors},\ }\href {https://doi.org/10.1103/RevModPhys.83.1057} {\bibfield  {journal} {\bibinfo  {journal} {Rev. Mod. Phys.}\ }\textbf {\bibinfo {volume} {83}},\ \bibinfo {pages} {1057} (\bibinfo {year} {2011})}\BibitemShut {NoStop}%
\bibitem [{\citenamefont {Khanikaev}\ \emph {et~al.}(2012)\citenamefont {Khanikaev}, \citenamefont {Mousavi}, \citenamefont {Tse}, \citenamefont {Kargarian}, \citenamefont {MacDonald},\ and\ \citenamefont {Shvets}}]{Khanikaev2012}%
  \BibitemOpen
  \bibfield  {author} {\bibinfo {author} {\bibfnamefont {A.~B.}\ \bibnamefont {Khanikaev}}, \bibinfo {author} {\bibfnamefont {S.~H.}\ \bibnamefont {Mousavi}}, \bibinfo {author} {\bibfnamefont {W.-K.}\ \bibnamefont {Tse}}, \bibinfo {author} {\bibfnamefont {M.}~\bibnamefont {Kargarian}}, \bibinfo {author} {\bibfnamefont {A.~H.}\ \bibnamefont {MacDonald}},\ and\ \bibinfo {author} {\bibfnamefont {G.}~\bibnamefont {Shvets}},\ }\bibfield  {title} {\bibinfo {title} {Photonic topological insulators},\ }\href {https://doi.org/10.1038/nmat3520} {\bibfield  {journal} {\bibinfo  {journal} {Nat. Mater.}\ }\textbf {\bibinfo {volume} {12}},\ \bibinfo {pages} {233} (\bibinfo {year} {2012})}\BibitemShut {NoStop}%
\bibitem [{\citenamefont {Lu}\ \emph {et~al.}(2014)\citenamefont {Lu}, \citenamefont {Joannopoulos},\ and\ \citenamefont {Solja{\v{c}}i{\'{c}}}}]{Lu2014}%
  \BibitemOpen
  \bibfield  {author} {\bibinfo {author} {\bibfnamefont {L.}~\bibnamefont {Lu}}, \bibinfo {author} {\bibfnamefont {J.~D.}\ \bibnamefont {Joannopoulos}},\ and\ \bibinfo {author} {\bibfnamefont {M.}~\bibnamefont {Solja{\v{c}}i{\'{c}}}},\ }\bibfield  {title} {\bibinfo {title} {Topological photonics},\ }\href {https://doi.org/10.1038/nphoton.2014.248} {\bibfield  {journal} {\bibinfo  {journal} {Nat. photonics}\ }\textbf {\bibinfo {volume} {8}},\ \bibinfo {pages} {821} (\bibinfo {year} {2014})}\BibitemShut {NoStop}%
\bibitem [{\citenamefont {Ozawa}\ \emph {et~al.}(2019)\citenamefont {Ozawa}, \citenamefont {Price}, \citenamefont {Amo}, \citenamefont {Goldman}, \citenamefont {Hafezi}, \citenamefont {Lu}, \citenamefont {Rechtsman}, \citenamefont {Schuster}, \citenamefont {Simon}, \citenamefont {Zilberberg},\ and\ \citenamefont {Carusotto}}]{RevModPhys.91.015006}%
  \BibitemOpen
  \bibfield  {author} {\bibinfo {author} {\bibfnamefont {T.}~\bibnamefont {Ozawa}}, \bibinfo {author} {\bibfnamefont {H.~M.}\ \bibnamefont {Price}}, \bibinfo {author} {\bibfnamefont {A.}~\bibnamefont {Amo}}, \bibinfo {author} {\bibfnamefont {N.}~\bibnamefont {Goldman}}, \bibinfo {author} {\bibfnamefont {M.}~\bibnamefont {Hafezi}}, \bibinfo {author} {\bibfnamefont {L.}~\bibnamefont {Lu}}, \bibinfo {author} {\bibfnamefont {M.~C.}\ \bibnamefont {Rechtsman}}, \bibinfo {author} {\bibfnamefont {D.}~\bibnamefont {Schuster}}, \bibinfo {author} {\bibfnamefont {J.}~\bibnamefont {Simon}}, \bibinfo {author} {\bibfnamefont {O.}~\bibnamefont {Zilberberg}},\ and\ \bibinfo {author} {\bibfnamefont {I.}~\bibnamefont {Carusotto}},\ }\bibfield  {title} {\bibinfo {title} {Topological photonics},\ }\href {https://doi.org/10.1103/RevModPhys.91.015006} {\bibfield  {journal} {\bibinfo  {journal} {Rev. Mod. Phys.}\ }\textbf {\bibinfo {volume} {91}},\ \bibinfo {pages} {015006} (\bibinfo {year} {2019})}\BibitemShut {NoStop}%
\bibitem [{\citenamefont {Karzig}\ \emph {et~al.}(2015)\citenamefont {Karzig}, \citenamefont {Bardyn}, \citenamefont {Lindner},\ and\ \citenamefont {Refael}}]{PhysRevX.5.031001}%
  \BibitemOpen
  \bibfield  {author} {\bibinfo {author} {\bibfnamefont {T.}~\bibnamefont {Karzig}}, \bibinfo {author} {\bibfnamefont {C.-E.}\ \bibnamefont {Bardyn}}, \bibinfo {author} {\bibfnamefont {N.~H.}\ \bibnamefont {Lindner}},\ and\ \bibinfo {author} {\bibfnamefont {G.}~\bibnamefont {Refael}},\ }\bibfield  {title} {\bibinfo {title} {Topological polaritons},\ }\href {https://doi.org/10.1103/PhysRevX.5.031001} {\bibfield  {journal} {\bibinfo  {journal} {Phys. Rev. X}\ }\textbf {\bibinfo {volume} {5}},\ \bibinfo {pages} {031001} (\bibinfo {year} {2015})}\BibitemShut {NoStop}%
\bibitem [{\citenamefont {Nalitov}\ \emph {et~al.}(2015)\citenamefont {Nalitov}, \citenamefont {Solnyshkov},\ and\ \citenamefont {Malpuech}}]{PhysRevLett.114.116401}%
  \BibitemOpen
  \bibfield  {author} {\bibinfo {author} {\bibfnamefont {A.~V.}\ \bibnamefont {Nalitov}}, \bibinfo {author} {\bibfnamefont {D.~D.}\ \bibnamefont {Solnyshkov}},\ and\ \bibinfo {author} {\bibfnamefont {G.}~\bibnamefont {Malpuech}},\ }\bibfield  {title} {\bibinfo {title} {Polariton $\mathbb{Z}$ topological insulator},\ }\href {https://doi.org/10.1103/PhysRevLett.114.116401} {\bibfield  {journal} {\bibinfo  {journal} {Phys. Rev. Lett.}\ }\textbf {\bibinfo {volume} {114}},\ \bibinfo {pages} {116401} (\bibinfo {year} {2015})}\BibitemShut {NoStop}%
\bibitem [{\citenamefont {St-Jean}\ \emph {et~al.}(2017{\natexlab{a}})\citenamefont {St-Jean}, \citenamefont {Goblot}, \citenamefont {Galopin}, \citenamefont {Lema{\^\i}tre}, \citenamefont {Ozawa}, \citenamefont {Le~Gratiet}, \citenamefont {Sagnes}, \citenamefont {Bloch},\ and\ \citenamefont {Amo}}]{st2017lasing}%
  \BibitemOpen
  \bibfield  {author} {\bibinfo {author} {\bibfnamefont {P.}~\bibnamefont {St-Jean}}, \bibinfo {author} {\bibfnamefont {V.}~\bibnamefont {Goblot}}, \bibinfo {author} {\bibfnamefont {E.}~\bibnamefont {Galopin}}, \bibinfo {author} {\bibfnamefont {A.}~\bibnamefont {Lema{\^\i}tre}}, \bibinfo {author} {\bibfnamefont {T.}~\bibnamefont {Ozawa}}, \bibinfo {author} {\bibfnamefont {L.}~\bibnamefont {Le~Gratiet}}, \bibinfo {author} {\bibfnamefont {I.}~\bibnamefont {Sagnes}}, \bibinfo {author} {\bibfnamefont {J.}~\bibnamefont {Bloch}},\ and\ \bibinfo {author} {\bibfnamefont {A.}~\bibnamefont {Amo}},\ }\bibfield  {title} {\bibinfo {title} {Lasing in topological edge states of a one-dimensional lattice},\ }\href@noop {} {\bibfield  {journal} {\bibinfo  {journal} {Nat. Photonics}\ }\textbf {\bibinfo {volume} {11}},\ \bibinfo {pages} {651} (\bibinfo {year} {2017}{\natexlab{a}})}\BibitemShut {NoStop}%
\bibitem [{\citenamefont {Klembt}\ \emph {et~al.}(2018)\citenamefont {Klembt}, \citenamefont {Harder}, \citenamefont {Egorov}, \citenamefont {Winkler}, \citenamefont {Ge}, \citenamefont {Bandres}, \citenamefont {Emmerling}, \citenamefont {Worschech}, \citenamefont {Liew}, \citenamefont {Segev}, \citenamefont {Schneider},\ and\ \citenamefont {Höfling}}]{klembt2018exciton}%
  \BibitemOpen
  \bibfield  {author} {\bibinfo {author} {\bibfnamefont {S.}~\bibnamefont {Klembt}}, \bibinfo {author} {\bibfnamefont {T.~H.}\ \bibnamefont {Harder}}, \bibinfo {author} {\bibfnamefont {O.~A.}\ \bibnamefont {Egorov}}, \bibinfo {author} {\bibfnamefont {K.}~\bibnamefont {Winkler}}, \bibinfo {author} {\bibfnamefont {R.}~\bibnamefont {Ge}}, \bibinfo {author} {\bibfnamefont {M.~A.}\ \bibnamefont {Bandres}}, \bibinfo {author} {\bibfnamefont {M.}~\bibnamefont {Emmerling}}, \bibinfo {author} {\bibfnamefont {L.}~\bibnamefont {Worschech}}, \bibinfo {author} {\bibfnamefont {T.~C.~H.}\ \bibnamefont {Liew}}, \bibinfo {author} {\bibfnamefont {M.}~\bibnamefont {Segev}}, \bibinfo {author} {\bibfnamefont {C.}~\bibnamefont {Schneider}},\ and\ \bibinfo {author} {\bibfnamefont {S.}~\bibnamefont {Höfling}},\ }\bibfield  {title} {\bibinfo {title} {Exciton-polariton topological insulator},\ }\href@noop {} {\bibfield  {journal} {\bibinfo  {journal} {Nature}\ }\textbf {\bibinfo {volume} {562}},\ \bibinfo {pages} {552} (\bibinfo
  {year} {2018})}\BibitemShut {NoStop}%
\bibitem [{\citenamefont {Kartashov}\ and\ \citenamefont {Skryabin}(2019)}]{PhysRevLett.122.083902}%
  \BibitemOpen
  \bibfield  {author} {\bibinfo {author} {\bibfnamefont {Y.~V.}\ \bibnamefont {Kartashov}}\ and\ \bibinfo {author} {\bibfnamefont {D.~V.}\ \bibnamefont {Skryabin}},\ }\bibfield  {title} {\bibinfo {title} {Two-dimensional topological polariton laser},\ }\href {https://doi.org/10.1103/PhysRevLett.122.083902} {\bibfield  {journal} {\bibinfo  {journal} {Phys. Rev. Lett.}\ }\textbf {\bibinfo {volume} {122}},\ \bibinfo {pages} {083902} (\bibinfo {year} {2019})}\BibitemShut {NoStop}%
\bibitem [{\citenamefont {Baranov}\ \emph {et~al.}(2020)\citenamefont {Baranov}, \citenamefont {Munkhbat}, \citenamefont {Zhukova}, \citenamefont {Bisht}, \citenamefont {Canales}, \citenamefont {Rousseaux}, \citenamefont {Johansson}, \citenamefont {Antosiewicz},\ and\ \citenamefont {Shegai}}]{Baranov2020}%
  \BibitemOpen
  \bibfield  {author} {\bibinfo {author} {\bibfnamefont {D.~G.}\ \bibnamefont {Baranov}}, \bibinfo {author} {\bibfnamefont {B.}~\bibnamefont {Munkhbat}}, \bibinfo {author} {\bibfnamefont {E.}~\bibnamefont {Zhukova}}, \bibinfo {author} {\bibfnamefont {A.}~\bibnamefont {Bisht}}, \bibinfo {author} {\bibfnamefont {A.}~\bibnamefont {Canales}}, \bibinfo {author} {\bibfnamefont {B.}~\bibnamefont {Rousseaux}}, \bibinfo {author} {\bibfnamefont {G.}~\bibnamefont {Johansson}}, \bibinfo {author} {\bibfnamefont {T.~J.}\ \bibnamefont {Antosiewicz}},\ and\ \bibinfo {author} {\bibfnamefont {T.}~\bibnamefont {Shegai}},\ }\bibfield  {title} {\bibinfo {title} {Ultrastrong coupling between nanoparticle plasmons and cavity photons at ambient conditions},\ }\href@noop {} {\bibfield  {journal} {\bibinfo  {journal} {Nat. Commun.}\ }\textbf {\bibinfo {volume} {11}} (\bibinfo {year} {2020})}\BibitemShut {NoStop}%
\bibitem [{\citenamefont {Liu}\ \emph {et~al.}(2020)\citenamefont {Liu}, \citenamefont {Ji}, \citenamefont {Wang}, \citenamefont {Modi}, \citenamefont {Hwang}, \citenamefont {Zheng}, \citenamefont {Sorger}, \citenamefont {Pan},\ and\ \citenamefont {Agarwal}}]{Liu_2020}%
  \BibitemOpen
  \bibfield  {author} {\bibinfo {author} {\bibfnamefont {W.}~\bibnamefont {Liu}}, \bibinfo {author} {\bibfnamefont {Z.}~\bibnamefont {Ji}}, \bibinfo {author} {\bibfnamefont {Y.}~\bibnamefont {Wang}}, \bibinfo {author} {\bibfnamefont {G.}~\bibnamefont {Modi}}, \bibinfo {author} {\bibfnamefont {M.}~\bibnamefont {Hwang}}, \bibinfo {author} {\bibfnamefont {B.}~\bibnamefont {Zheng}}, \bibinfo {author} {\bibfnamefont {V.~J.}\ \bibnamefont {Sorger}}, \bibinfo {author} {\bibfnamefont {A.}~\bibnamefont {Pan}},\ and\ \bibinfo {author} {\bibfnamefont {R.}~\bibnamefont {Agarwal}},\ }\bibfield  {title} {\bibinfo {title} {Generation of helical topological exciton-polaritons},\ }\href@noop {} {\bibfield  {journal} {\bibinfo  {journal} {Science}\ }\textbf {\bibinfo {volume} {370}},\ \bibinfo {pages} {600–604} (\bibinfo {year} {2020})}\BibitemShut {NoStop}%
\bibitem [{\citenamefont {Wu}\ \emph {et~al.}(2023)\citenamefont {Wu}, \citenamefont {Ghosh}, \citenamefont {Gan}, \citenamefont {Shi}, \citenamefont {Mandal}, \citenamefont {Sun}, \citenamefont {Zhang}, \citenamefont {Liew}, \citenamefont {Su},\ and\ \citenamefont {Xiong}}]{wu2023higher}%
  \BibitemOpen
  \bibfield  {author} {\bibinfo {author} {\bibfnamefont {J.}~\bibnamefont {Wu}}, \bibinfo {author} {\bibfnamefont {S.}~\bibnamefont {Ghosh}}, \bibinfo {author} {\bibfnamefont {Y.}~\bibnamefont {Gan}}, \bibinfo {author} {\bibfnamefont {Y.}~\bibnamefont {Shi}}, \bibinfo {author} {\bibfnamefont {S.}~\bibnamefont {Mandal}}, \bibinfo {author} {\bibfnamefont {H.}~\bibnamefont {Sun}}, \bibinfo {author} {\bibfnamefont {B.}~\bibnamefont {Zhang}}, \bibinfo {author} {\bibfnamefont {T.~C.}\ \bibnamefont {Liew}}, \bibinfo {author} {\bibfnamefont {R.}~\bibnamefont {Su}},\ and\ \bibinfo {author} {\bibfnamefont {Q.}~\bibnamefont {Xiong}},\ }\bibfield  {title} {\bibinfo {title} {Higher-order topological polariton corner state lasing},\ }\href@noop {} {\bibfield  {journal} {\bibinfo  {journal} {Sci. Adv.}\ }\textbf {\bibinfo {volume} {9}},\ \bibinfo {pages} {eadg4322} (\bibinfo {year} {2023})}\BibitemShut {NoStop}%
\bibitem [{\citenamefont {Hu}\ \emph {et~al.}(2023)\citenamefont {Hu}, \citenamefont {Chen}, \citenamefont {Teng}, \citenamefont {Yu}, \citenamefont {Xue}, \citenamefont {Chen}, \citenamefont {Xiao}, \citenamefont {Qu}, \citenamefont {Hu}, \citenamefont {Chen}, \citenamefont {Sun}, \citenamefont {Li}, \citenamefont {de~Abajo},\ and\ \citenamefont {Dai}}]{hu2023gate}%
  \BibitemOpen
  \bibfield  {author} {\bibinfo {author} {\bibfnamefont {H.}~\bibnamefont {Hu}}, \bibinfo {author} {\bibfnamefont {N.}~\bibnamefont {Chen}}, \bibinfo {author} {\bibfnamefont {H.}~\bibnamefont {Teng}}, \bibinfo {author} {\bibfnamefont {R.}~\bibnamefont {Yu}}, \bibinfo {author} {\bibfnamefont {M.}~\bibnamefont {Xue}}, \bibinfo {author} {\bibfnamefont {K.}~\bibnamefont {Chen}}, \bibinfo {author} {\bibfnamefont {Y.}~\bibnamefont {Xiao}}, \bibinfo {author} {\bibfnamefont {Y.}~\bibnamefont {Qu}}, \bibinfo {author} {\bibfnamefont {D.}~\bibnamefont {Hu}}, \bibinfo {author} {\bibfnamefont {J.}~\bibnamefont {Chen}}, \bibinfo {author} {\bibfnamefont {Z.}~\bibnamefont {Sun}}, \bibinfo {author} {\bibfnamefont {P.}~\bibnamefont {Li}}, \bibinfo {author} {\bibfnamefont {F.~J.~G.}\ \bibnamefont {de~Abajo}},\ and\ \bibinfo {author} {\bibfnamefont {Q.}~\bibnamefont {Dai}},\ }\bibfield  {title} {\bibinfo {title} {Gate-tunable negative refraction of mid-infrared polaritons},\ }\href@noop {} {\bibfield  {journal} {\bibinfo
  {journal} {Science}\ }\textbf {\bibinfo {volume} {379}},\ \bibinfo {pages} {558} (\bibinfo {year} {2023})}\BibitemShut {NoStop}%
\bibitem [{\citenamefont {Hassan}\ \emph {et~al.}(2019)\citenamefont {Hassan}, \citenamefont {Kunst}, \citenamefont {Moritz}, \citenamefont {Andler}, \citenamefont {Bergholtz},\ and\ \citenamefont {Bourennane}}]{ElHassan2019}%
  \BibitemOpen
  \bibfield  {author} {\bibinfo {author} {\bibfnamefont {A.~E.}\ \bibnamefont {Hassan}}, \bibinfo {author} {\bibfnamefont {F.~K.}\ \bibnamefont {Kunst}}, \bibinfo {author} {\bibfnamefont {A.}~\bibnamefont {Moritz}}, \bibinfo {author} {\bibfnamefont {G.}~\bibnamefont {Andler}}, \bibinfo {author} {\bibfnamefont {E.~J.}\ \bibnamefont {Bergholtz}},\ and\ \bibinfo {author} {\bibfnamefont {M.}~\bibnamefont {Bourennane}},\ }\bibfield  {title} {\bibinfo {title} {Corner states of light in photonic waveguides},\ }\href {https://doi.org/10.1038/s41566-019-0519-y} {\bibfield  {journal} {\bibinfo  {journal} {Nat. photonics}\ }\textbf {\bibinfo {volume} {13}},\ \bibinfo {pages} {697} (\bibinfo {year} {2019})}\BibitemShut {NoStop}%
\bibitem [{\citenamefont {Chen}\ \emph {et~al.}(2021)\citenamefont {Chen}, \citenamefont {Zhang}, \citenamefont {Chen}, \citenamefont {Yan}, \citenamefont {Xi}, \citenamefont {Chen},\ and\ \citenamefont {Yang}}]{chen2021photonic}%
  \BibitemOpen
  \bibfield  {author} {\bibinfo {author} {\bibfnamefont {Q.}~\bibnamefont {Chen}}, \bibinfo {author} {\bibfnamefont {L.}~\bibnamefont {Zhang}}, \bibinfo {author} {\bibfnamefont {F.}~\bibnamefont {Chen}}, \bibinfo {author} {\bibfnamefont {Q.}~\bibnamefont {Yan}}, \bibinfo {author} {\bibfnamefont {R.}~\bibnamefont {Xi}}, \bibinfo {author} {\bibfnamefont {H.}~\bibnamefont {Chen}},\ and\ \bibinfo {author} {\bibfnamefont {Y.}~\bibnamefont {Yang}},\ }\bibfield  {title} {\bibinfo {title} {Photonic topological valley-locked waveguides},\ }\href@noop {} {\bibfield  {journal} {\bibinfo  {journal} {ACS Photonics}\ }\textbf {\bibinfo {volume} {8}},\ \bibinfo {pages} {1400} (\bibinfo {year} {2021})}\BibitemShut {NoStop}%
\bibitem [{\citenamefont {Bahari}\ \emph {et~al.}(2017)\citenamefont {Bahari}, \citenamefont {Ndao}, \citenamefont {Vallini}, \citenamefont {Amili}, \citenamefont {Fainman},\ and\ \citenamefont {Kant{\'{e}}}}]{bahari2017nonreciprocal}%
  \BibitemOpen
  \bibfield  {author} {\bibinfo {author} {\bibfnamefont {B.}~\bibnamefont {Bahari}}, \bibinfo {author} {\bibfnamefont {A.}~\bibnamefont {Ndao}}, \bibinfo {author} {\bibfnamefont {F.}~\bibnamefont {Vallini}}, \bibinfo {author} {\bibfnamefont {A.~E.}\ \bibnamefont {Amili}}, \bibinfo {author} {\bibfnamefont {Y.}~\bibnamefont {Fainman}},\ and\ \bibinfo {author} {\bibfnamefont {B.}~\bibnamefont {Kant{\'{e}}}},\ }\bibfield  {title} {\bibinfo {title} {Nonreciprocal lasing in topological cavities of arbitrary geometries},\ }\href {https://doi.org/10.1126/science.aao4551} {\bibfield  {journal} {\bibinfo  {journal} {Science}\ }\textbf {\bibinfo {volume} {358}},\ \bibinfo {pages} {636} (\bibinfo {year} {2017})}\BibitemShut {NoStop}%
\bibitem [{\citenamefont {Barczyk}\ \emph {et~al.}(2022)\citenamefont {Barczyk}, \citenamefont {Parappurath}, \citenamefont {Arora}, \citenamefont {Bauer}, \citenamefont {Kuipers},\ and\ \citenamefont {Verhagen}}]{barczyk2022interplay}%
  \BibitemOpen
  \bibfield  {author} {\bibinfo {author} {\bibfnamefont {R.}~\bibnamefont {Barczyk}}, \bibinfo {author} {\bibfnamefont {N.}~\bibnamefont {Parappurath}}, \bibinfo {author} {\bibfnamefont {S.}~\bibnamefont {Arora}}, \bibinfo {author} {\bibfnamefont {T.}~\bibnamefont {Bauer}}, \bibinfo {author} {\bibfnamefont {L.}~\bibnamefont {Kuipers}},\ and\ \bibinfo {author} {\bibfnamefont {E.}~\bibnamefont {Verhagen}},\ }\bibfield  {title} {\bibinfo {title} {Interplay of leakage radiation and protection in topological photonic crystal cavities},\ }\href@noop {} {\bibfield  {journal} {\bibinfo  {journal} {Laser Photonics Rev.}\ }\textbf {\bibinfo {volume} {16}},\ \bibinfo {pages} {2200071} (\bibinfo {year} {2022})}\BibitemShut {NoStop}%
\bibitem [{\citenamefont {St-Jean}\ \emph {et~al.}(2017{\natexlab{b}})\citenamefont {St-Jean}, \citenamefont {Goblot}, \citenamefont {Galopin}, \citenamefont {Lema{\^{\i}}tre}, \citenamefont {Ozawa}, \citenamefont {Gratiet}, \citenamefont {Sagnes}, \citenamefont {Bloch},\ and\ \citenamefont {Amo}}]{StJean2017}%
  \BibitemOpen
  \bibfield  {author} {\bibinfo {author} {\bibfnamefont {P.}~\bibnamefont {St-Jean}}, \bibinfo {author} {\bibfnamefont {V.}~\bibnamefont {Goblot}}, \bibinfo {author} {\bibfnamefont {E.}~\bibnamefont {Galopin}}, \bibinfo {author} {\bibfnamefont {A.}~\bibnamefont {Lema{\^{\i}}tre}}, \bibinfo {author} {\bibfnamefont {T.}~\bibnamefont {Ozawa}}, \bibinfo {author} {\bibfnamefont {L.~L.}\ \bibnamefont {Gratiet}}, \bibinfo {author} {\bibfnamefont {I.}~\bibnamefont {Sagnes}}, \bibinfo {author} {\bibfnamefont {J.}~\bibnamefont {Bloch}},\ and\ \bibinfo {author} {\bibfnamefont {A.}~\bibnamefont {Amo}},\ }\bibfield  {title} {\bibinfo {title} {Lasing in topological edge states of a one-dimensional lattice},\ }\href {https://doi.org/10.1038/s41566-017-0006-2} {\bibfield  {journal} {\bibinfo  {journal} {Nat. Photonics}\ }\textbf {\bibinfo {volume} {11}},\ \bibinfo {pages} {651} (\bibinfo {year} {2017}{\natexlab{b}})}\BibitemShut {NoStop}%
\bibitem [{\citenamefont {Harari}\ \emph {et~al.}(2018)\citenamefont {Harari}, \citenamefont {Bandres}, \citenamefont {Lumer}, \citenamefont {Rechtsman}, \citenamefont {Chong}, \citenamefont {Khajavikhan}, \citenamefont {Christodoulides},\ and\ \citenamefont {Segev}}]{harari2018topological}%
  \BibitemOpen
  \bibfield  {author} {\bibinfo {author} {\bibfnamefont {G.}~\bibnamefont {Harari}}, \bibinfo {author} {\bibfnamefont {M.~A.}\ \bibnamefont {Bandres}}, \bibinfo {author} {\bibfnamefont {Y.}~\bibnamefont {Lumer}}, \bibinfo {author} {\bibfnamefont {M.~C.}\ \bibnamefont {Rechtsman}}, \bibinfo {author} {\bibfnamefont {Y.~D.}\ \bibnamefont {Chong}}, \bibinfo {author} {\bibfnamefont {M.}~\bibnamefont {Khajavikhan}}, \bibinfo {author} {\bibfnamefont {D.~N.}\ \bibnamefont {Christodoulides}},\ and\ \bibinfo {author} {\bibfnamefont {M.}~\bibnamefont {Segev}},\ }\bibfield  {title} {\bibinfo {title} {Topological insulator laser: Theory},\ }\href {https://doi.org/10.1126/science.aar4003} {\bibfield  {journal} {\bibinfo  {journal} {Science}\ }\textbf {\bibinfo {volume} {359}},\ \bibinfo {pages} {eaar4003} (\bibinfo {year} {2018})}\BibitemShut {NoStop}%
\bibitem [{\citenamefont {Bandres}\ \emph {et~al.}(2018)\citenamefont {Bandres}, \citenamefont {Wittek}, \citenamefont {Harari}, \citenamefont {Parto}, \citenamefont {Ren}, \citenamefont {Segev}, \citenamefont {Christodoulides},\ and\ \citenamefont {Khajavikhan}}]{bandres2018topological}%
  \BibitemOpen
  \bibfield  {author} {\bibinfo {author} {\bibfnamefont {M.~A.}\ \bibnamefont {Bandres}}, \bibinfo {author} {\bibfnamefont {S.}~\bibnamefont {Wittek}}, \bibinfo {author} {\bibfnamefont {G.}~\bibnamefont {Harari}}, \bibinfo {author} {\bibfnamefont {M.}~\bibnamefont {Parto}}, \bibinfo {author} {\bibfnamefont {J.}~\bibnamefont {Ren}}, \bibinfo {author} {\bibfnamefont {M.}~\bibnamefont {Segev}}, \bibinfo {author} {\bibfnamefont {D.~N.}\ \bibnamefont {Christodoulides}},\ and\ \bibinfo {author} {\bibfnamefont {M.}~\bibnamefont {Khajavikhan}},\ }\bibfield  {title} {\bibinfo {title} {Topological insulator laser: Experiments},\ }\href {https://doi.org/10.1126/science.aar4005} {\bibfield  {journal} {\bibinfo  {journal} {Science}\ }\textbf {\bibinfo {volume} {359}},\ \bibinfo {pages} {eaar4005} (\bibinfo {year} {2018})}\BibitemShut {NoStop}%
\bibitem [{\citenamefont {Yang}\ \emph {et~al.}(2022)\citenamefont {Yang}, \citenamefont {Li}, \citenamefont {Gao},\ and\ \citenamefont {Lu}}]{yang2022topological}%
  \BibitemOpen
  \bibfield  {author} {\bibinfo {author} {\bibfnamefont {L.}~\bibnamefont {Yang}}, \bibinfo {author} {\bibfnamefont {G.}~\bibnamefont {Li}}, \bibinfo {author} {\bibfnamefont {X.}~\bibnamefont {Gao}},\ and\ \bibinfo {author} {\bibfnamefont {L.}~\bibnamefont {Lu}},\ }\bibfield  {title} {\bibinfo {title} {Topological-cavity surface-emitting laser},\ }\href@noop {} {\bibfield  {journal} {\bibinfo  {journal} {Nat. Photonics}\ }\textbf {\bibinfo {volume} {16}},\ \bibinfo {pages} {279} (\bibinfo {year} {2022})}\BibitemShut {NoStop}%
\bibitem [{\citenamefont {Ningyuan}\ \emph {et~al.}(2015)\citenamefont {Ningyuan}, \citenamefont {Owens}, \citenamefont {Sommer}, \citenamefont {Schuster},\ and\ \citenamefont {Simon}}]{PhysRevX.5.021031}%
  \BibitemOpen
  \bibfield  {author} {\bibinfo {author} {\bibfnamefont {J.}~\bibnamefont {Ningyuan}}, \bibinfo {author} {\bibfnamefont {C.}~\bibnamefont {Owens}}, \bibinfo {author} {\bibfnamefont {A.}~\bibnamefont {Sommer}}, \bibinfo {author} {\bibfnamefont {D.}~\bibnamefont {Schuster}},\ and\ \bibinfo {author} {\bibfnamefont {J.}~\bibnamefont {Simon}},\ }\bibfield  {title} {\bibinfo {title} {Time- and site-resolved dynamics in a topological circuit},\ }\href {https://doi.org/10.1103/PhysRevX.5.021031} {\bibfield  {journal} {\bibinfo  {journal} {Phys. Rev. X}\ }\textbf {\bibinfo {volume} {5}},\ \bibinfo {pages} {021031} (\bibinfo {year} {2015})}\BibitemShut {NoStop}%
\bibitem [{\citenamefont {Ma}\ \emph {et~al.}(2019)\citenamefont {Ma}, \citenamefont {Xi},\ and\ \citenamefont {Sun}}]{ma2019topological}%
  \BibitemOpen
  \bibfield  {author} {\bibinfo {author} {\bibfnamefont {J.}~\bibnamefont {Ma}}, \bibinfo {author} {\bibfnamefont {X.}~\bibnamefont {Xi}},\ and\ \bibinfo {author} {\bibfnamefont {X.}~\bibnamefont {Sun}},\ }\bibfield  {title} {\bibinfo {title} {Topological photonic integrated circuits based on valley kink states},\ }\href@noop {} {\bibfield  {journal} {\bibinfo  {journal} {Laser Photonics Rev.}\ }\textbf {\bibinfo {volume} {13}},\ \bibinfo {pages} {1900087} (\bibinfo {year} {2019})}\BibitemShut {NoStop}%
\bibitem [{\citenamefont {Kruk}\ \emph {et~al.}(2019)\citenamefont {Kruk}, \citenamefont {Poddubny}, \citenamefont {Smirnova}, \citenamefont {Wang}, \citenamefont {Slobozhanyuk}, \citenamefont {Shorokhov}, \citenamefont {Kravchenko}, \citenamefont {Luther-Davies},\ and\ \citenamefont {Kivshar}}]{kruk2019nonlinear}%
  \BibitemOpen
  \bibfield  {author} {\bibinfo {author} {\bibfnamefont {S.}~\bibnamefont {Kruk}}, \bibinfo {author} {\bibfnamefont {A.}~\bibnamefont {Poddubny}}, \bibinfo {author} {\bibfnamefont {D.}~\bibnamefont {Smirnova}}, \bibinfo {author} {\bibfnamefont {L.}~\bibnamefont {Wang}}, \bibinfo {author} {\bibfnamefont {A.}~\bibnamefont {Slobozhanyuk}}, \bibinfo {author} {\bibfnamefont {A.}~\bibnamefont {Shorokhov}}, \bibinfo {author} {\bibfnamefont {I.}~\bibnamefont {Kravchenko}}, \bibinfo {author} {\bibfnamefont {B.}~\bibnamefont {Luther-Davies}},\ and\ \bibinfo {author} {\bibfnamefont {Y.}~\bibnamefont {Kivshar}},\ }\bibfield  {title} {\bibinfo {title} {Nonlinear light generation in topological nanostructures},\ }\href@noop {} {\bibfield  {journal} {\bibinfo  {journal} {Nat. Nanotechnol.}\ }\textbf {\bibinfo {volume} {14}},\ \bibinfo {pages} {126} (\bibinfo {year} {2019})}\BibitemShut {NoStop}%
\bibitem [{\citenamefont {Smirnova}\ \emph {et~al.}(2020)\citenamefont {Smirnova}, \citenamefont {Leykam}, \citenamefont {Chong},\ and\ \citenamefont {Kivshar}}]{Smirnova2020}%
  \BibitemOpen
  \bibfield  {author} {\bibinfo {author} {\bibfnamefont {D.}~\bibnamefont {Smirnova}}, \bibinfo {author} {\bibfnamefont {D.}~\bibnamefont {Leykam}}, \bibinfo {author} {\bibfnamefont {Y.}~\bibnamefont {Chong}},\ and\ \bibinfo {author} {\bibfnamefont {Y.}~\bibnamefont {Kivshar}},\ }\bibfield  {title} {\bibinfo {title} {Nonlinear topological photonics},\ }\href {https://doi.org/10.1063/1.5142397} {\bibfield  {journal} {\bibinfo  {journal} {Appl. Phys. Rev.}\ }\textbf {\bibinfo {volume} {7}},\ \bibinfo {pages} {021306} (\bibinfo {year} {2020})}\BibitemShut {NoStop}%
\bibitem [{\citenamefont {Midya}\ \emph {et~al.}(2018)\citenamefont {Midya}, \citenamefont {Zhao},\ and\ \citenamefont {Feng}}]{midya2018non}%
  \BibitemOpen
  \bibfield  {author} {\bibinfo {author} {\bibfnamefont {B.}~\bibnamefont {Midya}}, \bibinfo {author} {\bibfnamefont {H.}~\bibnamefont {Zhao}},\ and\ \bibinfo {author} {\bibfnamefont {L.}~\bibnamefont {Feng}},\ }\bibfield  {title} {\bibinfo {title} {Non-{H}ermitian photonics promises exceptional topology of light},\ }\href@noop {} {\bibfield  {journal} {\bibinfo  {journal} {Nat. Commun.}\ }\textbf {\bibinfo {volume} {9}},\ \bibinfo {pages} {2674} (\bibinfo {year} {2018})}\BibitemShut {NoStop}%
\bibitem [{\citenamefont {El-Ganainy}\ \emph {et~al.}(2018)\citenamefont {El-Ganainy}, \citenamefont {Makris}, \citenamefont {Khajavikhan}, \citenamefont {Musslimani}, \citenamefont {Rotter},\ and\ \citenamefont {Christodoulides}}]{ElGanainy2018}%
  \BibitemOpen
  \bibfield  {author} {\bibinfo {author} {\bibfnamefont {R.}~\bibnamefont {El-Ganainy}}, \bibinfo {author} {\bibfnamefont {K.~G.}\ \bibnamefont {Makris}}, \bibinfo {author} {\bibfnamefont {M.}~\bibnamefont {Khajavikhan}}, \bibinfo {author} {\bibfnamefont {Z.~H.}\ \bibnamefont {Musslimani}}, \bibinfo {author} {\bibfnamefont {S.}~\bibnamefont {Rotter}},\ and\ \bibinfo {author} {\bibfnamefont {D.~N.}\ \bibnamefont {Christodoulides}},\ }\bibfield  {title} {\bibinfo {title} {Non-{H}ermitian physics and {PT} symmetry},\ }\href {https://doi.org/10.1038/nphys4323} {\bibfield  {journal} {\bibinfo  {journal} {Nat. Phys.}\ }\textbf {\bibinfo {volume} {14}},\ \bibinfo {pages} {11} (\bibinfo {year} {2018})}\BibitemShut {NoStop}%
\bibitem [{\citenamefont {Parto}\ \emph {et~al.}(2020)\citenamefont {Parto}, \citenamefont {Liu}, \citenamefont {Bahari}, \citenamefont {Khajavikhan},\ and\ \citenamefont {Christodoulides}}]{parto2020non}%
  \BibitemOpen
  \bibfield  {author} {\bibinfo {author} {\bibfnamefont {M.}~\bibnamefont {Parto}}, \bibinfo {author} {\bibfnamefont {Y.~G.}\ \bibnamefont {Liu}}, \bibinfo {author} {\bibfnamefont {B.}~\bibnamefont {Bahari}}, \bibinfo {author} {\bibfnamefont {M.}~\bibnamefont {Khajavikhan}},\ and\ \bibinfo {author} {\bibfnamefont {D.~N.}\ \bibnamefont {Christodoulides}},\ }\bibfield  {title} {\bibinfo {title} {Non-{H}ermitian and topological photonics: optics at an exceptional point},\ }\href@noop {} {\bibfield  {journal} {\bibinfo  {journal} {Nanophotonics}\ }\textbf {\bibinfo {volume} {10}},\ \bibinfo {pages} {403} (\bibinfo {year} {2020})}\BibitemShut {NoStop}%
\bibitem [{\citenamefont {Zak}(1989)}]{PhysRevLett.62.2747}%
  \BibitemOpen
  \bibfield  {author} {\bibinfo {author} {\bibfnamefont {J.}~\bibnamefont {Zak}},\ }\bibfield  {title} {\bibinfo {title} {Berry's phase for energy bands in solids},\ }\href {https://doi.org/10.1103/PhysRevLett.62.2747} {\bibfield  {journal} {\bibinfo  {journal} {Phys. Rev. Lett.}\ }\textbf {\bibinfo {volume} {62}},\ \bibinfo {pages} {2747} (\bibinfo {year} {1989})}\BibitemShut {NoStop}%
\bibitem [{\citenamefont {Su}\ \emph {et~al.}(1979)\citenamefont {Su}, \citenamefont {Schrieffer},\ and\ \citenamefont {Heeger}}]{PhysRevLett.42.1698}%
  \BibitemOpen
  \bibfield  {author} {\bibinfo {author} {\bibfnamefont {W.~P.}\ \bibnamefont {Su}}, \bibinfo {author} {\bibfnamefont {J.~R.}\ \bibnamefont {Schrieffer}},\ and\ \bibinfo {author} {\bibfnamefont {A.~J.}\ \bibnamefont {Heeger}},\ }\bibfield  {title} {\bibinfo {title} {Solitons in polyacetylene},\ }\href {https://doi.org/10.1103/PhysRevLett.42.1698} {\bibfield  {journal} {\bibinfo  {journal} {Phys. Rev. Lett.}\ }\textbf {\bibinfo {volume} {42}},\ \bibinfo {pages} {1698} (\bibinfo {year} {1979})}\BibitemShut {NoStop}%
\bibitem [{\citenamefont {Haldane}\ and\ \citenamefont {Raghu}(2008)}]{PhysRevLett.100.013904}%
  \BibitemOpen
  \bibfield  {author} {\bibinfo {author} {\bibfnamefont {F.~D.~M.}\ \bibnamefont {Haldane}}\ and\ \bibinfo {author} {\bibfnamefont {S.}~\bibnamefont {Raghu}},\ }\bibfield  {title} {\bibinfo {title} {Possible realization of directional optical waveguides in photonic crystals with broken time-reversal symmetry},\ }\href {https://doi.org/10.1103/PhysRevLett.100.013904} {\bibfield  {journal} {\bibinfo  {journal} {Phys. Rev. Lett.}\ }\textbf {\bibinfo {volume} {100}},\ \bibinfo {pages} {013904} (\bibinfo {year} {2008})}\BibitemShut {NoStop}%
\bibitem [{\citenamefont {Wang}\ \emph {et~al.}(2009)\citenamefont {Wang}, \citenamefont {Chong}, \citenamefont {Joannopoulos},\ and\ \citenamefont {Solja{\v{c}}i{\'c}}}]{wang2009observation}%
  \BibitemOpen
  \bibfield  {author} {\bibinfo {author} {\bibfnamefont {Z.}~\bibnamefont {Wang}}, \bibinfo {author} {\bibfnamefont {Y.}~\bibnamefont {Chong}}, \bibinfo {author} {\bibfnamefont {J.~D.}\ \bibnamefont {Joannopoulos}},\ and\ \bibinfo {author} {\bibfnamefont {M.}~\bibnamefont {Solja{\v{c}}i{\'c}}},\ }\bibfield  {title} {\bibinfo {title} {Observation of unidirectional backscattering-immune topological electromagnetic states},\ }\href@noop {} {\bibfield  {journal} {\bibinfo  {journal} {Nature}\ }\textbf {\bibinfo {volume} {461}},\ \bibinfo {pages} {772} (\bibinfo {year} {2009})}\BibitemShut {NoStop}%
\bibitem [{\citenamefont {Raimond}\ \emph {et~al.}(2001)\citenamefont {Raimond}, \citenamefont {Brune},\ and\ \citenamefont {Haroche}}]{RevModPhys.73.565}%
  \BibitemOpen
  \bibfield  {author} {\bibinfo {author} {\bibfnamefont {J.~M.}\ \bibnamefont {Raimond}}, \bibinfo {author} {\bibfnamefont {M.}~\bibnamefont {Brune}},\ and\ \bibinfo {author} {\bibfnamefont {S.}~\bibnamefont {Haroche}},\ }\bibfield  {title} {\bibinfo {title} {Manipulating quantum entanglement with atoms and photons in a cavity},\ }\href {https://doi.org/10.1103/RevModPhys.73.565} {\bibfield  {journal} {\bibinfo  {journal} {Rev. Mod. Phys.}\ }\textbf {\bibinfo {volume} {73}},\ \bibinfo {pages} {565} (\bibinfo {year} {2001})}\BibitemShut {NoStop}%
\bibitem [{\citenamefont {Walther}\ \emph {et~al.}(2006)\citenamefont {Walther}, \citenamefont {Varcoe}, \citenamefont {Englert},\ and\ \citenamefont {Becker}}]{walther2006cavity}%
  \BibitemOpen
  \bibfield  {author} {\bibinfo {author} {\bibfnamefont {H.}~\bibnamefont {Walther}}, \bibinfo {author} {\bibfnamefont {B.~T.}\ \bibnamefont {Varcoe}}, \bibinfo {author} {\bibfnamefont {B.-G.}\ \bibnamefont {Englert}},\ and\ \bibinfo {author} {\bibfnamefont {T.}~\bibnamefont {Becker}},\ }\bibfield  {title} {\bibinfo {title} {Cavity quantum electrodynamics},\ }\href@noop {} {\bibfield  {journal} {\bibinfo  {journal} {Rep. Prog. Phys.}\ }\textbf {\bibinfo {volume} {69}},\ \bibinfo {pages} {1325} (\bibinfo {year} {2006})}\BibitemShut {NoStop}%
\bibitem [{\citenamefont {Mirhosseini}\ \emph {et~al.}(2019)\citenamefont {Mirhosseini}, \citenamefont {Kim}, \citenamefont {Zhang}, \citenamefont {Sipahigil}, \citenamefont {Dieterle}, \citenamefont {Keller}, \citenamefont {Asenjo-Garcia}, \citenamefont {Chang},\ and\ \citenamefont {Painter}}]{mirhosseini2019cavity}%
  \BibitemOpen
  \bibfield  {author} {\bibinfo {author} {\bibfnamefont {M.}~\bibnamefont {Mirhosseini}}, \bibinfo {author} {\bibfnamefont {E.}~\bibnamefont {Kim}}, \bibinfo {author} {\bibfnamefont {X.}~\bibnamefont {Zhang}}, \bibinfo {author} {\bibfnamefont {A.}~\bibnamefont {Sipahigil}}, \bibinfo {author} {\bibfnamefont {P.~B.}\ \bibnamefont {Dieterle}}, \bibinfo {author} {\bibfnamefont {A.~J.}\ \bibnamefont {Keller}}, \bibinfo {author} {\bibfnamefont {A.}~\bibnamefont {Asenjo-Garcia}}, \bibinfo {author} {\bibfnamefont {D.~E.}\ \bibnamefont {Chang}},\ and\ \bibinfo {author} {\bibfnamefont {O.}~\bibnamefont {Painter}},\ }\bibfield  {title} {\bibinfo {title} {Cavity quantum electrodynamics with atom-like mirrors},\ }\href@noop {} {\bibfield  {journal} {\bibinfo  {journal} {Nature}\ }\textbf {\bibinfo {volume} {569}},\ \bibinfo {pages} {692} (\bibinfo {year} {2019})}\BibitemShut {NoStop}%
\bibitem [{\citenamefont {Owens}\ \emph {et~al.}(2022)\citenamefont {Owens}, \citenamefont {Panetta}, \citenamefont {Saxberg}, \citenamefont {Roberts}, \citenamefont {Chakram}, \citenamefont {Ma}, \citenamefont {Vrajitoarea}, \citenamefont {Simon},\ and\ \citenamefont {Schuster}}]{owens2022chiral}%
  \BibitemOpen
  \bibfield  {author} {\bibinfo {author} {\bibfnamefont {J.~C.}\ \bibnamefont {Owens}}, \bibinfo {author} {\bibfnamefont {M.~G.}\ \bibnamefont {Panetta}}, \bibinfo {author} {\bibfnamefont {B.}~\bibnamefont {Saxberg}}, \bibinfo {author} {\bibfnamefont {G.}~\bibnamefont {Roberts}}, \bibinfo {author} {\bibfnamefont {S.}~\bibnamefont {Chakram}}, \bibinfo {author} {\bibfnamefont {R.}~\bibnamefont {Ma}}, \bibinfo {author} {\bibfnamefont {A.}~\bibnamefont {Vrajitoarea}}, \bibinfo {author} {\bibfnamefont {J.}~\bibnamefont {Simon}},\ and\ \bibinfo {author} {\bibfnamefont {D.~I.}\ \bibnamefont {Schuster}},\ }\bibfield  {title} {\bibinfo {title} {Chiral cavity quantum electrodynamics},\ }\href@noop {} {\bibfield  {journal} {\bibinfo  {journal} {Nat. Phys.}\ }\textbf {\bibinfo {volume} {18}},\ \bibinfo {pages} {1048} (\bibinfo {year} {2022})}\BibitemShut {NoStop}%
\bibitem [{\citenamefont {Lei}\ \emph {et~al.}(2023)\citenamefont {Lei}, \citenamefont {Fukumori}, \citenamefont {Rochman}, \citenamefont {Zhu}, \citenamefont {Endres}, \citenamefont {Choi},\ and\ \citenamefont {Faraon}}]{lei2023many}%
  \BibitemOpen
  \bibfield  {author} {\bibinfo {author} {\bibfnamefont {M.}~\bibnamefont {Lei}}, \bibinfo {author} {\bibfnamefont {R.}~\bibnamefont {Fukumori}}, \bibinfo {author} {\bibfnamefont {J.}~\bibnamefont {Rochman}}, \bibinfo {author} {\bibfnamefont {B.}~\bibnamefont {Zhu}}, \bibinfo {author} {\bibfnamefont {M.}~\bibnamefont {Endres}}, \bibinfo {author} {\bibfnamefont {J.}~\bibnamefont {Choi}},\ and\ \bibinfo {author} {\bibfnamefont {A.}~\bibnamefont {Faraon}},\ }\bibfield  {title} {\bibinfo {title} {Many-body cavity quantum electrodynamics with driven inhomogeneous emitters},\ }\href@noop {} {\bibfield  {journal} {\bibinfo  {journal} {Nature}\ }\textbf {\bibinfo {volume} {617}},\ \bibinfo {pages} {271–276} (\bibinfo {year} {2023})}\BibitemShut {NoStop}%
\bibitem [{\citenamefont {Wang}\ \emph {et~al.}(2018)\citenamefont {Wang}, \citenamefont {Zhang}, \citenamefont {Zhang}, \citenamefont {Li}, \citenamefont {Hu},\ and\ \citenamefont {You}}]{PhysRevLett.120.057202}%
  \BibitemOpen
  \bibfield  {author} {\bibinfo {author} {\bibfnamefont {Y.-P.}\ \bibnamefont {Wang}}, \bibinfo {author} {\bibfnamefont {G.-Q.}\ \bibnamefont {Zhang}}, \bibinfo {author} {\bibfnamefont {D.}~\bibnamefont {Zhang}}, \bibinfo {author} {\bibfnamefont {T.-F.}\ \bibnamefont {Li}}, \bibinfo {author} {\bibfnamefont {C.-M.}\ \bibnamefont {Hu}},\ and\ \bibinfo {author} {\bibfnamefont {J.~Q.}\ \bibnamefont {You}},\ }\bibfield  {title} {\bibinfo {title} {Bistability of cavity magnon polaritons},\ }\href {https://doi.org/10.1103/PhysRevLett.120.057202} {\bibfield  {journal} {\bibinfo  {journal} {Phys. Rev. Lett.}\ }\textbf {\bibinfo {volume} {120}},\ \bibinfo {pages} {057202} (\bibinfo {year} {2018})}\BibitemShut {NoStop}%
\bibitem [{\citenamefont {Shen}\ \emph {et~al.}(2021)\citenamefont {Shen}, \citenamefont {Wang}, \citenamefont {Li}, \citenamefont {Zhu}, \citenamefont {Agarwal},\ and\ \citenamefont {You}}]{PhysRevLett.127.183202}%
  \BibitemOpen
  \bibfield  {author} {\bibinfo {author} {\bibfnamefont {R.-C.}\ \bibnamefont {Shen}}, \bibinfo {author} {\bibfnamefont {Y.-P.}\ \bibnamefont {Wang}}, \bibinfo {author} {\bibfnamefont {J.}~\bibnamefont {Li}}, \bibinfo {author} {\bibfnamefont {S.-Y.}\ \bibnamefont {Zhu}}, \bibinfo {author} {\bibfnamefont {G.~S.}\ \bibnamefont {Agarwal}},\ and\ \bibinfo {author} {\bibfnamefont {J.~Q.}\ \bibnamefont {You}},\ }\bibfield  {title} {\bibinfo {title} {Long-time memory and ternary logic gate using a multistable cavity magnonic system},\ }\href {https://doi.org/10.1103/PhysRevLett.127.183202} {\bibfield  {journal} {\bibinfo  {journal} {Phys. Rev. Lett.}\ }\textbf {\bibinfo {volume} {127}},\ \bibinfo {pages} {183202} (\bibinfo {year} {2021})}\BibitemShut {NoStop}%
\bibitem [{\citenamefont {Shen}\ \emph {et~al.}(2022)\citenamefont {Shen}, \citenamefont {Li}, \citenamefont {Fan}, \citenamefont {Wang},\ and\ \citenamefont {You}}]{PhysRevLett.129.123601}%
  \BibitemOpen
  \bibfield  {author} {\bibinfo {author} {\bibfnamefont {R.-C.}\ \bibnamefont {Shen}}, \bibinfo {author} {\bibfnamefont {J.}~\bibnamefont {Li}}, \bibinfo {author} {\bibfnamefont {Z.-Y.}\ \bibnamefont {Fan}}, \bibinfo {author} {\bibfnamefont {Y.-P.}\ \bibnamefont {Wang}},\ and\ \bibinfo {author} {\bibfnamefont {J.~Q.}\ \bibnamefont {You}},\ }\bibfield  {title} {\bibinfo {title} {Mechanical bistability in kerr-modified cavity magnomechanics},\ }\href {https://doi.org/10.1103/PhysRevLett.129.123601} {\bibfield  {journal} {\bibinfo  {journal} {Phys. Rev. Lett.}\ }\textbf {\bibinfo {volume} {129}},\ \bibinfo {pages} {123601} (\bibinfo {year} {2022})}\BibitemShut {NoStop}%
\bibitem [{\citenamefont {Chanda}\ \emph {et~al.}(2011)\citenamefont {Chanda}, \citenamefont {Shigeta}, \citenamefont {Truong}, \citenamefont {Lui}, \citenamefont {Mihi}, \citenamefont {Schulmerich}, \citenamefont {Braun}, \citenamefont {Bhargava},\ and\ \citenamefont {Rogers}}]{chanda2011coupling}%
  \BibitemOpen
  \bibfield  {author} {\bibinfo {author} {\bibfnamefont {D.}~\bibnamefont {Chanda}}, \bibinfo {author} {\bibfnamefont {K.}~\bibnamefont {Shigeta}}, \bibinfo {author} {\bibfnamefont {T.}~\bibnamefont {Truong}}, \bibinfo {author} {\bibfnamefont {E.}~\bibnamefont {Lui}}, \bibinfo {author} {\bibfnamefont {A.}~\bibnamefont {Mihi}}, \bibinfo {author} {\bibfnamefont {M.}~\bibnamefont {Schulmerich}}, \bibinfo {author} {\bibfnamefont {P.~V.}\ \bibnamefont {Braun}}, \bibinfo {author} {\bibfnamefont {R.}~\bibnamefont {Bhargava}},\ and\ \bibinfo {author} {\bibfnamefont {J.~A.}\ \bibnamefont {Rogers}},\ }\bibfield  {title} {\bibinfo {title} {Coupling of plasmonic and optical cavity modes in quasi-three-dimensional plasmonic crystals},\ }\href@noop {} {\bibfield  {journal} {\bibinfo  {journal} {Nat. commun.}\ }\textbf {\bibinfo {volume} {2}},\ \bibinfo {pages} {479} (\bibinfo {year} {2011})}\BibitemShut {NoStop}%
\bibitem [{\citenamefont {Hugall}\ \emph {et~al.}(2018)\citenamefont {Hugall}, \citenamefont {Singh},\ and\ \citenamefont {van Hulst}}]{hugall2018plasmonic}%
  \BibitemOpen
  \bibfield  {author} {\bibinfo {author} {\bibfnamefont {J.~T.}\ \bibnamefont {Hugall}}, \bibinfo {author} {\bibfnamefont {A.}~\bibnamefont {Singh}},\ and\ \bibinfo {author} {\bibfnamefont {N.~F.}\ \bibnamefont {van Hulst}},\ }\bibfield  {title} {\bibinfo {title} {Plasmonic cavity coupling},\ }\href@noop {} {\bibfield  {journal} {\bibinfo  {journal} {Acs Photonics}\ }\textbf {\bibinfo {volume} {5}},\ \bibinfo {pages} {43} (\bibinfo {year} {2018})}\BibitemShut {NoStop}%
\bibitem [{\citenamefont {Garcia-Vidal}\ \emph {et~al.}(2022)\citenamefont {Garcia-Vidal}, \citenamefont {Fern\'andez-Dom\'{\i}nguez}, \citenamefont {Martin-Moreno}, \citenamefont {Zhang}, \citenamefont {Tang}, \citenamefont {Peng},\ and\ \citenamefont {Cui}}]{RevModPhys.94.025004}%
  \BibitemOpen
  \bibfield  {author} {\bibinfo {author} {\bibfnamefont {F.~J.}\ \bibnamefont {Garcia-Vidal}}, \bibinfo {author} {\bibfnamefont {A.~I.}\ \bibnamefont {Fern\'andez-Dom\'{\i}nguez}}, \bibinfo {author} {\bibfnamefont {L.}~\bibnamefont {Martin-Moreno}}, \bibinfo {author} {\bibfnamefont {H.~C.}\ \bibnamefont {Zhang}}, \bibinfo {author} {\bibfnamefont {W.}~\bibnamefont {Tang}}, \bibinfo {author} {\bibfnamefont {R.}~\bibnamefont {Peng}},\ and\ \bibinfo {author} {\bibfnamefont {T.~J.}\ \bibnamefont {Cui}},\ }\bibfield  {title} {\bibinfo {title} {Spoof surface plasmon photonics},\ }\href {https://doi.org/10.1103/RevModPhys.94.025004} {\bibfield  {journal} {\bibinfo  {journal} {Rev. Mod. Phys.}\ }\textbf {\bibinfo {volume} {94}},\ \bibinfo {pages} {025004} (\bibinfo {year} {2022})}\BibitemShut {NoStop}%
\bibitem [{\citenamefont {Downing}\ \emph {et~al.}(2019)\citenamefont {Downing}, \citenamefont {Sturges}, \citenamefont {Weick}, \citenamefont {Stobi\ifmmode~\acute{n}\else \'{n}\fi{}ska},\ and\ \citenamefont {Mart\'{\i}n-Moreno}}]{PhysRevLett.123.217401}%
  \BibitemOpen
  \bibfield  {author} {\bibinfo {author} {\bibfnamefont {C.~A.}\ \bibnamefont {Downing}}, \bibinfo {author} {\bibfnamefont {T.~J.}\ \bibnamefont {Sturges}}, \bibinfo {author} {\bibfnamefont {G.}~\bibnamefont {Weick}}, \bibinfo {author} {\bibfnamefont {M.}~\bibnamefont {Stobi\ifmmode~\acute{n}\else \'{n}\fi{}ska}},\ and\ \bibinfo {author} {\bibfnamefont {L.}~\bibnamefont {Mart\'{\i}n-Moreno}},\ }\bibfield  {title} {\bibinfo {title} {Topological phases of polaritons in a cavity waveguide},\ }\href {https://link.aps.org/doi/10.1103/PhysRevLett.123.217401} {\bibfield  {journal} {\bibinfo  {journal} {Phys. Rev. Lett.}\ }\textbf {\bibinfo {volume} {123}},\ \bibinfo {pages} {217401} (\bibinfo {year} {2019})}\BibitemShut {NoStop}%
\bibitem [{\citenamefont {Mann}\ \emph {et~al.}(2018)\citenamefont {Mann}, \citenamefont {Sturges}, \citenamefont {Weick}, \citenamefont {Barnes},\ and\ \citenamefont {Mariani}}]{mann2018manipulating}%
  \BibitemOpen
  \bibfield  {author} {\bibinfo {author} {\bibfnamefont {C.-R.}\ \bibnamefont {Mann}}, \bibinfo {author} {\bibfnamefont {T.~J.}\ \bibnamefont {Sturges}}, \bibinfo {author} {\bibfnamefont {G.}~\bibnamefont {Weick}}, \bibinfo {author} {\bibfnamefont {W.~L.}\ \bibnamefont {Barnes}},\ and\ \bibinfo {author} {\bibfnamefont {E.}~\bibnamefont {Mariani}},\ }\bibfield  {title} {\bibinfo {title} {Manipulating type-{I} and type-{II} {D}irac polaritons in cavity-embedded honeycomb metasurfaces},\ }\href@noop {} {\bibfield  {journal} {\bibinfo  {journal} {Nat. Commun.}\ }\textbf {\bibinfo {volume} {9}},\ \bibinfo {pages} {2194} (\bibinfo {year} {2018})}\BibitemShut {NoStop}%
\bibitem [{\citenamefont {Mann}\ \emph {et~al.}(2020)\citenamefont {Mann}, \citenamefont {Horsley},\ and\ \citenamefont {Mariani}}]{mann2020tunable}%
  \BibitemOpen
  \bibfield  {author} {\bibinfo {author} {\bibfnamefont {C.-R.}\ \bibnamefont {Mann}}, \bibinfo {author} {\bibfnamefont {S.~A.}\ \bibnamefont {Horsley}},\ and\ \bibinfo {author} {\bibfnamefont {E.}~\bibnamefont {Mariani}},\ }\bibfield  {title} {\bibinfo {title} {Tunable pseudo-magnetic fields for polaritons in strained metasurfaces},\ }\href@noop {} {\bibfield  {journal} {\bibinfo  {journal} {Nat. Photonics}\ }\textbf {\bibinfo {volume} {14}},\ \bibinfo {pages} {669} (\bibinfo {year} {2020})}\BibitemShut {NoStop}%
\bibitem [{\citenamefont {Downing}\ and\ \citenamefont {Mart{\'\i}n-Moreno}(2020)}]{downing2020polaritonic}%
  \BibitemOpen
  \bibfield  {author} {\bibinfo {author} {\bibfnamefont {C.~A.}\ \bibnamefont {Downing}}\ and\ \bibinfo {author} {\bibfnamefont {L.}~\bibnamefont {Mart{\'\i}n-Moreno}},\ }\bibfield  {title} {\bibinfo {title} {Polaritonic {T}amm states induced by cavity photons},\ }\href@noop {} {\bibfield  {journal} {\bibinfo  {journal} {Nanophotonics}\ }\textbf {\bibinfo {volume} {10}},\ \bibinfo {pages} {513} (\bibinfo {year} {2020})}\BibitemShut {NoStop}%
\bibitem [{\citenamefont {Xiao}\ \emph {et~al.}(2017)\citenamefont {Xiao}, \citenamefont {Ma}, \citenamefont {Zhang},\ and\ \citenamefont {Chan}}]{PhysRevLett.118.166803}%
  \BibitemOpen
  \bibfield  {author} {\bibinfo {author} {\bibfnamefont {Y.-X.}\ \bibnamefont {Xiao}}, \bibinfo {author} {\bibfnamefont {G.}~\bibnamefont {Ma}}, \bibinfo {author} {\bibfnamefont {Z.-Q.}\ \bibnamefont {Zhang}},\ and\ \bibinfo {author} {\bibfnamefont {C.~T.}\ \bibnamefont {Chan}},\ }\bibfield  {title} {\bibinfo {title} {Topological subspace-induced bound state in the continuum},\ }\href {https://doi.org/10.1103/PhysRevLett.118.166803} {\bibfield  {journal} {\bibinfo  {journal} {Phys. Rev. Lett.}\ }\textbf {\bibinfo {volume} {118}},\ \bibinfo {pages} {166803} (\bibinfo {year} {2017})}\BibitemShut {NoStop}%
\bibitem [{\citenamefont {Sheremet}\ \emph {et~al.}(2023)\citenamefont {Sheremet}, \citenamefont {Petrov}, \citenamefont {Iorsh}, \citenamefont {Poshakinskiy},\ and\ \citenamefont {Poddubny}}]{RevModPhys.95.015002}%
  \BibitemOpen
  \bibfield  {author} {\bibinfo {author} {\bibfnamefont {A.~S.}\ \bibnamefont {Sheremet}}, \bibinfo {author} {\bibfnamefont {M.~I.}\ \bibnamefont {Petrov}}, \bibinfo {author} {\bibfnamefont {I.~V.}\ \bibnamefont {Iorsh}}, \bibinfo {author} {\bibfnamefont {A.~V.}\ \bibnamefont {Poshakinskiy}},\ and\ \bibinfo {author} {\bibfnamefont {A.~N.}\ \bibnamefont {Poddubny}},\ }\bibfield  {title} {\bibinfo {title} {Waveguide quantum electrodynamics: Collective radiance and photon-photon correlations},\ }\href {https://doi.org/10.1103/RevModPhys.95.015002} {\bibfield  {journal} {\bibinfo  {journal} {Rev. Mod. Phys.}\ }\textbf {\bibinfo {volume} {95}},\ \bibinfo {pages} {015002} (\bibinfo {year} {2023})}\BibitemShut {NoStop}%
\bibitem [{\citenamefont {Sturges}\ \emph {et~al.}(2020)\citenamefont {Sturges}, \citenamefont {Rep{\"a}n}, \citenamefont {Downing}, \citenamefont {Rockstuhl},\ and\ \citenamefont {Stobi{\'n}ska}}]{Sturges_2020}%
  \BibitemOpen
  \bibfield  {author} {\bibinfo {author} {\bibfnamefont {T.~J.}\ \bibnamefont {Sturges}}, \bibinfo {author} {\bibfnamefont {T.}~\bibnamefont {Rep{\"a}n}}, \bibinfo {author} {\bibfnamefont {C.~A.}\ \bibnamefont {Downing}}, \bibinfo {author} {\bibfnamefont {C.}~\bibnamefont {Rockstuhl}},\ and\ \bibinfo {author} {\bibfnamefont {M.}~\bibnamefont {Stobi{\'n}ska}},\ }\bibfield  {title} {\bibinfo {title} {Extreme renormalisations of dimer eigenmodes by strong light--matter coupling},\ }\href@noop {} {\bibfield  {journal} {\bibinfo  {journal} {New J. Phys.}\ }\textbf {\bibinfo {volume} {22}},\ \bibinfo {pages} {103001} (\bibinfo {year} {2020})}\BibitemShut {NoStop}%
\bibitem [{\citenamefont {Downing}\ and\ \citenamefont {Weick}(2017)}]{PhysRevB.95.125426}%
  \BibitemOpen
  \bibfield  {author} {\bibinfo {author} {\bibfnamefont {C.~A.}\ \bibnamefont {Downing}}\ and\ \bibinfo {author} {\bibfnamefont {G.}~\bibnamefont {Weick}},\ }\bibfield  {title} {\bibinfo {title} {Topological collective plasmons in bipartite chains of metallic nanoparticles},\ }\href {https://link.aps.org/doi/10.1103/PhysRevB.95.125426} {\bibfield  {journal} {\bibinfo  {journal} {Phys. Rev. B}\ }\textbf {\bibinfo {volume} {95}},\ \bibinfo {pages} {125426} (\bibinfo {year} {2017})}\BibitemShut {NoStop}%
\bibitem [{\citenamefont {Xiao}\ \emph {et~al.}(2014)\citenamefont {Xiao}, \citenamefont {Zhang},\ and\ \citenamefont {Chan}}]{PhysRevX.4.021017}%
  \BibitemOpen
  \bibfield  {author} {\bibinfo {author} {\bibfnamefont {M.}~\bibnamefont {Xiao}}, \bibinfo {author} {\bibfnamefont {Z.~Q.}\ \bibnamefont {Zhang}},\ and\ \bibinfo {author} {\bibfnamefont {C.~T.}\ \bibnamefont {Chan}},\ }\bibfield  {title} {\bibinfo {title} {Surface impedance and bulk band geometric phases in one-dimensional systems},\ }\href {https://doi.org/10.1103/PhysRevX.4.021017} {\bibfield  {journal} {\bibinfo  {journal} {Phys. Rev. X}\ }\textbf {\bibinfo {volume} {4}},\ \bibinfo {pages} {021017} (\bibinfo {year} {2014})}\BibitemShut {NoStop}%
\bibitem [{\citenamefont {Xiao}\ \emph {et~al.}(2015)\citenamefont {Xiao}, \citenamefont {Ma}, \citenamefont {Yang}, \citenamefont {Sheng}, \citenamefont {Zhang},\ and\ \citenamefont {Chan}}]{xiao2015geometric}%
  \BibitemOpen
  \bibfield  {author} {\bibinfo {author} {\bibfnamefont {M.}~\bibnamefont {Xiao}}, \bibinfo {author} {\bibfnamefont {G.}~\bibnamefont {Ma}}, \bibinfo {author} {\bibfnamefont {Z.}~\bibnamefont {Yang}}, \bibinfo {author} {\bibfnamefont {P.}~\bibnamefont {Sheng}}, \bibinfo {author} {\bibfnamefont {Z.}~\bibnamefont {Zhang}},\ and\ \bibinfo {author} {\bibfnamefont {C.~T.}\ \bibnamefont {Chan}},\ }\bibfield  {title} {\bibinfo {title} {Geometric phase and band inversion in periodic acoustic systems},\ }\href@noop {} {\bibfield  {journal} {\bibinfo  {journal} {Nat. Phys.}\ }\textbf {\bibinfo {volume} {11}},\ \bibinfo {pages} {240} (\bibinfo {year} {2015})}\BibitemShut {NoStop}%
\bibitem [{\citenamefont {Li}\ \emph {et~al.}(2022)\citenamefont {Li}, \citenamefont {Du}, \citenamefont {Zhang}, \citenamefont {Li}, \citenamefont {Fan}, \citenamefont {Zhang},\ and\ \citenamefont {Qiu}}]{PhysRevLett.128.116803}%
  \BibitemOpen
  \bibfield  {author} {\bibinfo {author} {\bibfnamefont {T.}~\bibnamefont {Li}}, \bibinfo {author} {\bibfnamefont {J.}~\bibnamefont {Du}}, \bibinfo {author} {\bibfnamefont {Q.}~\bibnamefont {Zhang}}, \bibinfo {author} {\bibfnamefont {Y.}~\bibnamefont {Li}}, \bibinfo {author} {\bibfnamefont {X.}~\bibnamefont {Fan}}, \bibinfo {author} {\bibfnamefont {F.}~\bibnamefont {Zhang}},\ and\ \bibinfo {author} {\bibfnamefont {C.}~\bibnamefont {Qiu}},\ }\bibfield  {title} {\bibinfo {title} {Acoustic {M}\"obius insulators from projective symmetry},\ }\href {https://doi.org/10.1103/PhysRevLett.128.116803} {\bibfield  {journal} {\bibinfo  {journal} {Phys. Rev. Lett.}\ }\textbf {\bibinfo {volume} {128}},\ \bibinfo {pages} {116803} (\bibinfo {year} {2022})}\BibitemShut {NoStop}%
\bibitem [{\citenamefont {P\'erez-Gonz\'alez}\ \emph {et~al.}(2019)\citenamefont {P\'erez-Gonz\'alez}, \citenamefont {Bello}, \citenamefont {G\'omez-Le\'on},\ and\ \citenamefont {Platero}}]{PhysRevB.99.035146}%
  \BibitemOpen
  \bibfield  {author} {\bibinfo {author} {\bibfnamefont {B.}~\bibnamefont {P\'erez-Gonz\'alez}}, \bibinfo {author} {\bibfnamefont {M.}~\bibnamefont {Bello}}, \bibinfo {author} {\bibfnamefont {A.}~\bibnamefont {G\'omez-Le\'on}},\ and\ \bibinfo {author} {\bibfnamefont {G.}~\bibnamefont {Platero}},\ }\bibfield  {title} {\bibinfo {title} {Interplay between long-range hopping and disorder in topological systems},\ }\href {https://doi.org/10.1103/PhysRevB.99.035146} {\bibfield  {journal} {\bibinfo  {journal} {Phys. Rev. B}\ }\textbf {\bibinfo {volume} {99}},\ \bibinfo {pages} {035146} (\bibinfo {year} {2019})}\BibitemShut {NoStop}%
\end{thebibliography}
\end{document}